\documentclass[aps,english,preprint,nofootinbib,preprintnumbers]{revtex4-2}
\usepackage{mathrsfs}
\usepackage{hyperref}
\hypersetup{
    colorlinks=true,
    linkcolor=blue,
    filecolor=magenta,      
    urlcolor=cyan,
    pdfpagemode=FullScreen,
    }
\usepackage{graphicx}
\usepackage{amsmath, amssymb,bbm}
\usepackage{babel}
\usepackage{color}
\usepackage{slashed}
\usepackage[T1]{fontenc}
\usepackage{titlesec}
\usepackage{empheq}
\def\beqn{\begin{eqnarray}}
\def\eeqn{\end{eqnarray}}
\def\barr{\begin{array}}
\def\earr{\end{array}}
\def\btab{\begin{tabular}}
\def\etab{\end{tabular}}
\def\bite{\begin{itemize}}
\def\eite{\end{itemize}}
\def\bcen{\begin{center}}
\def\ecen{\end{center}}

\newcommand{\pnu}{|\vec{p}_\nu'|}

\begin{document}

\title{Pseudo-neutrino versus recoil formalism for 4-body phase space and applications to nuclear decay}

\author{Chien-Yeah Seng$^{1,2}$}

\affiliation{$^{1}$Facility for Rare Isotope Beams, Michigan State University, East Lansing, MI 48824, USA}
\affiliation{$^{2}$Department of Physics, University of Washington,
	Seattle, WA 98195-1560, USA}

\date{\today}

\begin{abstract}
	
It is well-known that the traditional treatment of radiative corrections that utilizes the ``true'' neutrino momentum $\vec{p}_\nu$ in the differential decay rate formula could lead to a $\sim \alpha/\pi$ systematic error in certain observables due to the mistreatment of 4-body kinematics. I investigate the theory structure of one of the proposed solutions, the ``$\nu'$-formalism'', in the non-recoil limit appropriate for neutron and nuclear beta decays. I derive an elegant master formula for the 4-body phase space and use it to re-analyze the spectrum-dependent ``outer'' radiative corrections to the beta decay of a polarized spin-half nucleus; a complete set of analytic expressions is provided for readers to straightforwardly obtain the final numerical results. I compare it to the ``recoil formalism'' where the energy of the recoil nucleus is fixed.

\end{abstract}

\maketitle
\newpage


\section{Introduction}

Neutron and nuclear beta decays provide an excellent avenue for precision tests of the Standard Model (SM)~\cite{Brodeur:2023eul,Alarcon:2023gfu}. Apart from the decay lifetime that determines the Cabibbo-Kobayashi-Maskawa (CKM) matrix $V_{ud}$, the study of the beta decay spectrum and various correlations constructed from spin and momentum vectors set stringent constraints on coupling strengths induced by physics beyond the Standard Model (BSM). The first comprehensive analysis of various terms in the beta decay differential rate in terms of the parameters in the most general Lee-Yang Lagrangian~\cite{Lee:1956qn} was done by Jackson, Treiman and Wyld ~\cite{Jackson:1957zz,Jackson:1957auh}. Taking neutron decay as an example, the well-known tree-level differential decay rate of a polarized neutron (with unpolarized final states) reads:
\begin{equation}
\left(\frac{d\Gamma}{dE_ed\Omega_e d\Omega_\nu}\right)_\text{tree}\propto 1+a\vec{\beta}\cdot\hat{p}_\nu+b\frac{m_e}{E_e}+\hat{s}_n\cdot\left[A\vec{\beta}+B\hat{p}_\nu+D\vec{\beta}\times\hat{p}_\nu\right]~,\label{eq:Gammatree}
\end{equation}
where $\hat{p}_\nu$ is the unit neutrino momentum vector, $\vec{\beta}=\vec{p}_e/E_e=\beta \hat{p}_e$ is the electron velocity vector, and $\hat{s}_n$ is the neutron unit polarization vector. The SM provides well-defined predictions for the correlation coefficients $a$, $b$, $A$ and $B$ and $D$;
experimental confirmation of any deviation from such predictions will be an unambiguous signal of BSM physics.   

As the present experimental precision of beta decay observables have reached $10^{-4}$, one needs to carefully account for the higher-order SM corrections to the tree-level expression; the most prominent ones are recoil effects and the radiative corrections (RC). The latter consists of the Fermi function~\cite{Fermi:1934hr} and ``outer corrections'' that are functions of $E_e$ and the angles, and ``inner corrections'' that only serve to renormalize weak coupling constants; the first two are more important for the determination of correlation coefficients. 

Pioneering works by Sirlin~\cite{Sirlin:1967zza}, Shann~\cite{Shann:1971fz} and Garcia-Maya~\cite{Garcia:1978bq} established the $\mathcal{O}(\alpha)$ outer corrections to the polarized neutron differential rate, which takes a very elegant form:
\begin{equation}
\delta\left(\frac{d\Gamma}{dE_ed\Omega_ed\Omega_\nu}\right)_\text{outer}\propto \frac{\alpha}{2\pi}\left\{\delta_1+a\vec{\beta}\cdot\hat{p}_\nu(\delta_1+\delta_2)+\hat{s}_n\cdot\left[A\vec{\beta}(\delta_1+\delta_2)+B\hat{p}_\nu\delta_1\right]\right\}~,\label{eq:tradhandwavy}
\end{equation}  
where $\delta_1$, $\delta_2$ are known functions of $E_e$. This equation serves as an important theory input for the study of the beta spectrum and the beta asymmetry parameter $A$. On the other hand, for observables such as the neutrino-electron correlation coefficient $a$ and the neutrino asymmetry parameter $B$, it was noticed long ago~\cite{Toth:1984ei} that the representation above, which I call the neutrino ($\nu$)-formalism, contains a serious drawback as it depends on the neutrino momentum $\vec{p}_\nu$ which is not directly measured. In a three-body decay $\phi_i\rightarrow \phi_fe\nu$, it could be deduced from the momentum conservation $\vec{p}_\nu=-\vec{p}_f-\vec{p}_e$, where both $\vec{p}_f$ and $\vec{p}_e$ are measurable quantities. However, this relation breaks down in the bremsstrahlung process where an extra real photon is emitted. in fact, since in most experiments the neutrino and the extra photon(s) are not directly detected, there is no way to determine the neutrino solid angle $\Omega_\nu$ using the momenta of $\phi_f$ and $e$. Putting it in other words, using the $\nu$-formalism for the differential rate together with the 3-body momentum conservation relations could lead to a theoretical error of the order $\alpha/\pi$ for the $\hat{p}_\nu$-dependent observables due to the mistreatment of the 4-body kinematics.  

Several alternative approaches were proposed to overcome the drawback in Eq.\eqref{eq:tradhandwavy}. An important example is the ``recoil formalism'', namely to fix the momenta $\vec{p}_e$ and $\vec{p}_f$ (instead of $\vec{p}_e$ and $\vec{p}_\nu$) in the calculation; in this way one may choose the variables as $E_e,E_f$ or $E_e,\cos\theta_{ef}\equiv\hat{p}_e\cdot \hat{p}_f$. This is analogous to the treatment of RC in semileptonic kaon decays~\cite{Ginsberg:1966zz,Ginsberg:1968pz,Ginsberg:1969jh,Ginsberg:1970vy}. The recoil formalism was used to study decays of unpolarized~\cite{Toth:1984er,Gluck:1992tg} and polarized~\cite{Gluck:1989sf,Gluck:1992qy,Gluck:1998ogp} baryons. The complicated bremsstrahlung phase space integrals were performed either semi-analytically~\cite{Gluck:1992tg,Gluck:1992qy} or numerically~\cite{Toth:1984er,Gluck:1993hh,Gluck:1994sw,Gluck:1997km}. Recently, the need to apply the correct recoil formalism (instead of the $\nu$-formalism) to the decay of free neutrons was reiterated~\cite{Gluck:2022ogz}, and subsequent re-analysis was performed to the aCORN~\cite{Wietfeldt:2023mdb} and aSPECT~\cite{Beck:2023hnt} data.  

Another approach is the pseudo-neutrino ($\nu'$)-formalism. It \textit{defines} a ``pseudo-neutrino'' momentum $\vec{p}_\nu'\equiv -\vec{p}_e-\vec{p}_f$ which is a direct experimental observable; this definition holds with or without extra photons. Subsequently, one expresses the differential rate in terms of $E_e,\Omega_e,\Omega_\nu'$ instead of $E_e,\Omega_e,\Omega_\nu$. This approach possesses several advantageous comparing to the recoil formalism: 
\begin{itemize}
	\item It is the most natural generalization to the traditional $\nu$-formalism and has the closest resemblance to the latter, which makes cross-checking easier. It is also the most natural formalism to describe decays of polarized nuclei. 
	\item It is much more natural when combined with non-recoil approximation (unlike the recoil formalism where the variation of $E_f$ cannot be discussed without involving recoil effects). It is sometimes beneficial to study both the kinematical dynamical recoil corrections simultaneously within a unified framework, e.g. using methods like the heavy baryon expansion~\cite{Jenkins:1990jv}, to avoid possible double-counting. So it is desirable to split recoil effects completely from RC, and the $\nu'$-formalism is best suited for this purpose.
	\item The analytic expressions of the phase space integrals in the $\nu'$-formalism are usually much more elegant, and only a minimal amount of numerical integration is required. 
\end{itemize}
Existing literature that discussed the $\nu'$-formalism includes Refs.\cite{Toth:1988fw,Gluck:1989sf,Gluck:1992qy,Gluck:1992tg}, with applications to decays of unpolarized and polarized neutron and hyperons. Despite their comprehensiveness, I failed to identify a complete treatment of the outer RC to the differential rate of a polarized parent nucleus in the non-recoil limit (appropriate for neutron and nuclear decays) with the $\nu'$-formalism which, in principle, should display a similar degree of theory elegance as Eq.\eqref{eq:tradhandwavy}. Also, analytic expressions of many important intermediate quantities (e.g. expansion coefficients and results of phase-space integrals) are not explicitly shown, which makes their results not so straightforwardly applicable in numerical analysis; furthermore, one also finds that some numerical results in these earlier papers are not completely accurate. In the meantime, while many recent computer packages are capable to produce reliable numerical results~\cite{Gluck:2022ogz}, they are practically a black box to outsiders.

The purpose of this paper is to provide a transparent, elegant and self-contained theoretical description 
of the $\nu'$-formalism in the non-recoil limit. As a simple application, I study the outer RC to the differential rate of the $\beta^{\pm}$ decay of a polarized spin-1/2 nucleus, but the final result is equally applicable to spin-0 systems. 
Therefore, this analysis covers a wide range of interesting beta decay processes including semileptonic pion decay, free neutron decay, superallowed $0^+\rightarrow 0^+$ decays~\cite{Hardy:2020qwl}, spin-half nuclear mirror transitions~\cite{NaviliatCuncic:2008xt} (e.g. $^{19}$Ne$\rightarrow$$^{19}$F), etc. I provide the full set of analytic expressions of all intermediate functions needed in the analysis, so that the quoted numerical results are fully transparent and can be easily reproduced by any interested reader, without the need of complicated numerical packages. For completeness, I also provide a similar set of analytic formula for unpolarized nuclei in the recoil formalism.  

\section{3-body phase space}

I start by briefly reviewing the total rate formula of a 3-body decay $\phi_i(p_i)\rightarrow \phi_f(p_f)+e(p_e)+\nu (p_\nu)$\footnote{The term ``electron'' throughout this paper may refer to an electron or a positron depending on the decay mode, and similar for the neutrino.}:
\begin{equation}
\Gamma_3=\int d\Pi_3|\mathcal{M}|_\text{3-body}^2
\end{equation}
with the 3-body phase space:
\begin{equation}
\int d\Pi_3=\frac{1}{2m_i}\int\frac{d^3p_f}{(2\pi)^32E_f}\frac{d^3p_e}{(2\pi)^32E_e}\frac{d^3p_\nu}{(2\pi)^32E_\nu}(2\pi)^4\delta^{(4)}(p_i-p_f-p_e-p_\nu)~.\label{eq:Pi3}
\end{equation}
One is interested in beta decay processes where $m_i,m_f\gg m_i-m_f$. The evaluation of this phase space in the $\nu$-formalism consists of first integrating out $\vec{p}_f$ using the spatial delta function. The remaining temporal delta function reads:
\begin{equation}
\delta(m_i-E_f-E_e-E_\nu)\approx \delta (E_m-E_e-E_\nu)
\end{equation}
which only set bounds on the energy variables but not the angles; here I have taken the non-recoil approximation, $E_f\approx m_f$ and $E_m\equiv (m_i^2-m_f^2+m_e^2)/(2m_i)\approx m_i-m_f$ is the electron end-point energy. Of course, in practice one needs to account for recoil corrections to the tree-level 3-body rate for precision, but that is not the focus of this paper. This delta function is then used to integrate out the neutrino energy, which leaves us with:
\begin{equation}
\int d\Pi_3\approx \frac{1}{512\pi^5m_im_f}\int d\Omega_ed\Omega_\nu \int_{m_e}^{E_m}dE_e|\vec{p}_e|E_\nu~,\label{eq:Pi3nu}
\end{equation}  
where $E_\nu=E_m-E_e$; the upper limit of $E_e$ is set by the temporal delta function: $E_m-E_e=E_\nu\geq 0$. 
Eq.\eqref{eq:Pi3nu} is the underlying reason why one utilized $E_e$, $\Omega_e$ and $\Omega_\nu$ as free variables in the traditional analysis of beta decays. 

To bypass the conceptual deficit in the $\nu$-formalism that $\Omega_\nu$ is not a practical observable, one may turn to the $\nu'$-formalism by defining the pseudo-neutrino momentum:   
\begin{equation}
p_\nu'\equiv p_i-p_f-p_e=(E_\nu',\vec{p}_\nu')\label{eq:pnup}
\end{equation}
which is an experimental observable, because both $e$ and $\phi_f$ can be directly detected. In the non-recoil limit, $E_\nu'=E_m-E_e$ is only a function of electron energy. In this formalism, I change the integration over $\vec{p}_f$ to that over $\vec{p}_\nu'$: 
\begin{equation}
\int d^3p_f d^3p_e=\int d^3p_\nu' d^3p_e~,
\end{equation}
and rewrite Eq.\eqref{eq:Pi3} in the non-recoil limit as:
\begin{equation}
\int d\Pi_3\approx\frac{1}{(2\pi)^68m_im_f}\int d^3p_\nu'\frac{d^3p_e}{E_e}\frac{d^3p_\nu}{(2\pi)^32E_\nu}(2\pi)^4\delta^{(4)}(p_\nu'-p_\nu)~,
\end{equation}
and integrate out $\vec{p}_\nu$, instead of $\vec{p}_f$, using the spatial delta function. The remaining temporal delta function reads $\delta (E_\nu'-\pnu)$, which is then used to integrate out $\pnu$. This yields:
\begin{equation}
\int d\Pi_3\approx \frac{1}{512\pi^5m_im_f}\int d\Omega_ed\Omega_\nu' \int_{m_e}^{E_m}dE_e|\vec{p}_e|E_\nu'~.\label{eq:Pi3nuprime}
\end{equation}  
This is identical to Eq.\eqref{eq:Pi3nu} upon replacing $E_\nu'\rightarrow E_\nu$ and  $\Omega_\nu'\rightarrow \Omega_\nu$, which do not change the physics since $p_\nu'=p_\nu$ in 3-body decay. The real distinction between the two formalisms only shows up when extra photons are emitted. 

To facilitate the discussion later, I also provide the representation of the 3-body phase space in the recoil formalism, i.e. using $E_e$ and $E_f$ as variables. It is most convenient to adopt the notations used in semileptonic decays of kaons~\cite{Ginsberg:1966zz,Ginsberg:1968pz,Ginsberg:1969jh,Ginsberg:1970vy}, where the differential decay rate is expressed in terms of the following dimensionless Lorentz scalars:
\begin{equation}
x\equiv \frac{p_\nu^{\prime 2}}{m_i^2}~,~y\equiv \frac{2p_i\cdot p_e}{m_i^2}~,~z\equiv \frac{2p_i\cdot p_f}{m_i^2}~,
\end{equation}
with $y=2E_e/m_i$ and $z=2E_f/m_i$ energy parameters in the parent's rest frame. In 3-body decay, $p_\nu^{\prime 2}=p_\nu^2=0$ assuming massless neutrino, and I obtain the \textit{exact} relation:
\begin{equation}
\int d\Pi_3=\frac{m_i}{256\pi^3}\int_{\mathcal{D}_3}dydz~,
\end{equation}
where the integration region $\mathcal{D}_3$ can be represented in two equivalent ways:
\begin{eqnarray}
a(y)-b(y)<z<a(y)+b(y)~,&&2\sqrt{r_e}<y<1+r_e-r_f\nonumber\\
a(y)=\frac{(2-y)(1+r_f+r_e-y)}{2(1+r_e-y)}~,&&b(y)=\frac{\sqrt{y^2-4r_e}(1+r_e-r_f-y)}{2(1+r_e-y)}~,
\end{eqnarray}
or
\begin{eqnarray}
c(z)-d(z)<y<c(z)+d(z)~,&&2\sqrt{r_f}<z<1+r_f-r_e\nonumber\\
c(z)=\frac{(2-z)(1+r_e+r_f-z)}{2(1+r_f-z)}~,&&d(z)=\frac{\sqrt{z^2-4r_f}(1+r_f-r_e-z)}{2(1+r_f-z)}~,
\end{eqnarray}
with $r_f\equiv m_f^2/m_i^2$, $r_e\equiv m_e^2/m_i^2$.

\section{4-body phase space}

The contribution from the bremsstrahlung process $\phi_i(p_i)\rightarrow \phi_f(p_f)+e(p_e)+\nu (p_\nu)+\gamma(k)$ to the total decay rate is:
\begin{equation}
\Gamma_4=\int d\Pi_4|\mathcal{M}|^2_\text{4-body}~,\label{eq:Gamma4}
\end{equation}
where the 4-body phase space reads:
\begin{equation}
\int d\Pi_4=\frac{1}{2m_i}\int\frac{d^3p_f}{(2\pi)^32E_f}\frac{d^3p_e}{(2\pi)^32E_e}\frac{d^3p_\nu}{(2\pi)^32E_\nu}\frac{d^3k}{(2\pi)^32E_k}(2\pi)^4\delta^{(4)}(p_i-p_f-p_e-p_\nu-k)~.
\end{equation}

Let us again evaluate the phase space first in the traditional $\nu$-formalism. I integrate out $\vec{p}_f$ using the spatial delta function, which leaves the temporal part as:
\begin{equation}
\delta (m_i-E_f-E_e-E_\nu-E_k)\approx \delta(E_m-E_e-E_\nu-E_k)~.
\end{equation}
Here the non-recoil approximation can be safely taken, because in neutron/nuclear beta decays a recoil correction on top of a RC gives a $\sim(\alpha/\pi)(m_i-m_f)/m_i\lesssim 10^{-5}$ correction to the total decay rate, which exceeds the experimental precision and can be neglected. The remaining delta function
is then used to integrate out $E_\nu$, which gives:
\begin{equation}
\int d\Pi_4\approx\frac{1}{512\pi^5m_im_f}\int d\Omega_ed\Omega_\nu\int_{m_e}^{E_m}dE_e|\vec{p}_e|\int_{E_k<E_m-E_e}\frac{d^3k}{(2\pi)^32E_k}E_\nu~,\label{eq:Pi4nu}
\end{equation}
where $E_\nu=E_m-E_e-E_k$. The unobserved photon momentum $\vec{k}$ is then integrated out, leaving $E_e$, $\Omega_e$ and $\Omega_\nu$ as variables in the differential decay rate formula. 

As I stressed in the introduction, since in Eq.\eqref{eq:Pi4nu} the unobserved photon momentum is integrated out, further applying the 3-body momentum conservation $p_\nu=p_i-p_f-p_e$ to the 4-body phase space in the $\nu$-formalism leads to an error. This issue is non-existent in the recoil formalism, since the variables $\{x,y,z\}$ are experimentally measurable. In this formalism, one can derive the following \textit{exact} formula of the 4-body phase space~\cite{Seng:2021wcf}:
\begin{eqnarray}
\int d\Pi_4&=&\frac{m_i^3}{512\pi^4}\left\{\int_{\mathcal{D}_3}dydz\int_0^{\alpha_+(y,z)}dx+\int_{\mathcal{D}_{4-3}}dydz\int_{\alpha_-(y,z)}^{\alpha_+(y,z)}dx\right\}\int\frac{d^3k}{(2\pi)^32E_k}\frac{d^3p_\nu}{(2\pi)^32E_\nu}\nonumber\\
&&\times (2\pi)^4\delta^{(4)}(p_\nu'-p_\nu-k)~,\label{eq:phaseexact}
\end{eqnarray}
where
\begin{equation}
\alpha_\pm (y,z)\equiv 1-y-z+r_f+r_e+\frac{yz}{2}\pm\frac{1}{2}\sqrt{y^2-4r_e}\sqrt{z^2-4r_f}~,
\end{equation}
and the region $\mathcal{D}_{4-3}$ can also be expressed in two equivalent ways:
\begin{equation}
2\sqrt{r_f}<z<a(y)-b(y)~,~2\sqrt{r_e}<y<1-\sqrt{r_f}+\frac{r_e}{1-\sqrt{r_f}}
\end{equation}
or
\begin{equation}
2\sqrt{r_e}<y<c(z)-d(z)~,~2\sqrt{r_f}<z<1-\sqrt{r_e}+\frac{r_f}{1-\sqrt{r_e}}~.
\end{equation}
This formula is extremely useful, because it turns out the $p_\nu$- and $k$-integration with respect to the squared amplitude is always analytically doable~\cite{Ginsberg:1969jh}, so one is left with at most one single numerical integration over $x$, if the differential rate is expressed as $d\Gamma/dydz$. The recoil formalism has also been applied to baryon decays~\cite{Toth:1984er,Gluck:1992tg,Gluck:1989sf,Gluck:1992qy,Gluck:1998ogp}; in particular, the non-overlapping regions  $\mathcal{D}_3$ and $\mathcal{D}_{4-3}$ in Eq.\eqref{eq:phaseexact} are just what known as the ``in'' and ``out'' region respectively in Ref.\cite{Gluck:2022ogz}.   
However, since $\{y,z\}$ are all energy and not angular variables, the connection to the traditional representation~\eqref{eq:Gammatree}, \eqref{eq:tradhandwavy} is not straightforward.

To overcome the aforementioned shortcomings, I re-evaluate the 4-body phase space in the $\nu'$-formalism, keeping close analogy to the  3-body case as well as the kaon representation, in order to retain the advantages in both methods. Notice that the definition of the pseudo-neutrino momentum $p_\nu'$ in Eq.\eqref{eq:pnup} remains unchanged in 4-body, and the relation $E_\nu'=E_m-E_e$ still holds. I first write:
\begin{equation}
\int d^3p_f\frac{d^3p_e}{E_e}=\int d^3p_\nu'\frac{d^3p_e}{E_e}=\int d\Omega_ed\Omega_\nu'\int dE_e|\vec{p}_e|\int d\pnu \pnu^2~,
\end{equation}
and deduce the boundaries of each variable. First, the boundaries of $E_e$, $\Omega_e$ and $\Omega_\nu'$ must be exactly the same as in the 3-body case; an easy way to understand this is that it must be so to ensure the exact cancellation of the infrared (IR)-divergence between the 3-body (one-loop) and 4-body (tree-level) contributions to the differential decay rate $d\Gamma/dE_ed\Omega_ed\Omega_\nu'$~\cite{Bloch:1937pw,Yennie:1961ad,Kinoshita:1958ru,Kinoshita:1962ur,Lee:1964is}. The boundaries of $\pnu$ are determined by the delta function:
\begin{equation}
p_\nu^{\prime 2}=E_\nu^{\prime 2}-\pnu^2=(p_\nu+k)^2=(E_\nu+E_k)^2-|\vec{p}_\nu+\vec{k}|^2\geq 0~, 
\end{equation}
so one has $\pnu\leq E_\nu'$, with the equal sign occurring when $\vec{p}_\nu\parallel\vec{k}$. Meanwhile, for any given $\vec{p}_e$, a $\vec{p}_\nu'=\vec{0}$ 
configuration can always be achieved by setting $\vec{p}_f=-\vec{p}_e$ (which does not affect the energy $E_f\approx m_f$ in the non-recoil limit), with the remaining energy of the system carried away by a $\nu \gamma$-pair with equal and opposite momenta. So, the lower and upper limits of the $\pnu$-integration are 0 and $E_\nu'$ respectively. Thus, I obtain the $\nu'$-representation of the 4-body phase space in the non-recoil limit: 
\begin{eqnarray}
\int d\Pi_4&\approx&\frac{1}{512\pi^6m_im_f}\int d\Omega_ed\Omega_{\nu}'\int_{m_e}^{E_m}dE_e|\vec{p}_e|\int_0^{E_\nu'}d\pnu\pnu^2\int\frac{d^3k}{(2\pi)^32E_k}\frac{d^3p_\nu}{(2\pi)^32E_\nu}\nonumber\\
&&\times (2\pi)^4\delta^{(4)}(p_\nu'-p_\nu-k)~.\label{eq:phaseapprox}
\end{eqnarray}

Eq.\eqref{eq:phaseapprox} is the first central result of this work. It possesses the following advantages:
\begin{enumerate}
	\item The differential rate is expressible in terms of the experimentally-measurable energy and angular variables $E_e$, $\Omega_e$ and $\Omega_\nu'$, which solves the problem in the traditional $\nu$-formalism.
	\item The mathematical structure of Eq.\eqref{eq:phaseapprox} is much more elegant than the recoil formalism, Eq.\eqref{eq:phaseexact}. For instance, here one does not need to separate the so-called ``in'' and ``out'' region as two different terms, and the boundaries are also much simpler.  
	\item The full $p_\nu$- and $k$-integrations are retained as in the kaon formalism. This allows one to directly make use of the known analytic formula of such integrations in the latter. 
\end{enumerate}
Finally, one can easily check the correctness of Eq.\eqref{eq:phaseapprox} by comparing it to Eq.\eqref{eq:phaseexact}. For instance, after removing the $p_\nu$- and $k$-measures and the delta function, I evaluate the two expressions numerically by substituting the neutron decay parameters, and find that their difference is at the level 0.1\% which is the typical size of a  
recoil correction  $\sim(m_n-m_p)/m_n$. This shows that Eq.\eqref{eq:phaseapprox} is indeed the correct 4-body phase space formula in the non-recoil limit.

\section{Radiative corrections to nuclear decay in $\nu'$-formalism}

As an important application of the $\nu'$-formalism, I study the so-called ``outer RC'' in a $J_i=J_f=1/2$ decay of a polarized parent nucleus (with unmeasured final spins) at the $\mathcal{O}(\alpha/\pi)$ level. This includes well-known one-loop corrections and the one-photon bremsstrahlung process. The former  does not distinguish between the $\nu$- and $\nu'$-method, so I focus on the latter. 

I start by writing down the hadronic and leptonic charged weak current operator\footnote{In this work I adopt the experimentalists' convention, $g_A<0$.}:
\begin{equation}
J_{\text{had}}^{\mu}=\left\{ \begin{array}{ccc}
\bar{n}\gamma^{\mu}(g_{V}+g_{A}\gamma_{5})p & , & \beta^{+}\\
\bar{p}\gamma^{\mu}(g_{V}+g_{A}\gamma_{5})n & , & \beta^{-}
\end{array}\right.~,~J_{\text{lep}}^{\mu}=\left\{ \begin{array}{ccc}
\bar{\nu}\gamma^{\mu}(1-\gamma_{5})e & , & \beta^{+}\\
\bar{e}\gamma^{\mu}(1-\gamma_{5})\nu & , & \beta^{-}
\end{array}\right.~.
\end{equation}
The Fermi (F) and Gamow-Teller (GT) matrix element of $J_\text{had}^\mu$ are defined through:
\begin{equation}
\langle f|\int d^3x J_\text{had}^0(\vec{x})|i\rangle=g_V M_\text{F}\mathbbm{1}~,~\langle f|\int d^3x \vec{J}_\text{had}(\vec{x})|i\rangle=\frac{g_A}{\sqrt{3}}M_\text{GT}\vec{\sigma}~.
\end{equation}  
With this one can define a ``generalized'' axial-to-vector ratio:
\begin{equation}
\lambda\equiv \frac{M_\text{GT}}{\sqrt{3}M_\text{F}}\frac{g_A}{g_V}~,
\end{equation}
which reduces to $g_A/g_V$ for neutron decay ($M_\text{F}=1$ and $M_\text{GT}=\sqrt{3}$). Taking the non-recoil limit at the amplitude level, which corresponds to setting $p_f\approx p_i\equiv p=(m,\vec{0})$ with $m=m_i$ for simplicity, I can write down the tree-level squared amplitude as:
\begin{equation}
|\mathcal{M}|_\text{tree}^2\approx 16G_V^2m^2E_eE_\nu'\left\{1+3\lambda^2+(1-\lambda^2)\vec{\beta}\cdot\hat{p}_\nu'+\hat{s}\cdot\left[2\lambda(\lambda\zeta-1)\vec{\beta}-2\lambda(\lambda\zeta+1)\hat{p}_\nu'\right]\right\}~,\label{eq:Mtree2}
\end{equation}
where $\zeta=+1(-1)$ for $\beta^+(\beta^-)$-decay, $G_V\equiv G_FV_{ud}g_V M_F$ is the effective weak vector coupling, and $\hat{s}$ is the unit vector of the parent's polarization. Plugging it into Eq.\eqref{eq:Pi3nuprime} gives the tree-level differential rate in the non-recoil limit (taking $m^2\approx m_i m_f$):
\begin{eqnarray}
\left(\frac{d\Gamma}{dE_ed\Omega_ed\Omega_\nu'}\right)_\text{tree}&\approx &\frac{G_V^2}{(2\pi)^5}|\vec{p}_e|E_eE_\nu^{\prime 2}\Bigl\{1+3\lambda^2+(1-\lambda^2)\vec{\beta}\cdot\hat{p}_\nu'\nonumber\\
&&+\hat{s}\cdot\left[2\lambda(\lambda\zeta-1)\vec{\beta}-2\lambda(\lambda\zeta+1)\hat{p}_\nu'\right]\Bigr\}~.
\end{eqnarray}

\subsection{\label{sec:Brsq}The bremsstrahlung squared amplitude}

\begin{figure}
	\begin{centering}
		\includegraphics[scale=0.8]{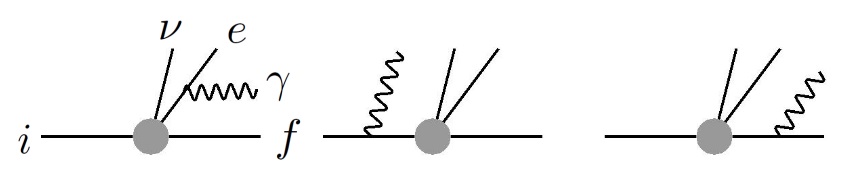}
		\par\end{centering}
	\caption{\label{fig:brem} Bremsstrahlung diagrams.  }
\end{figure}

I now write down the bremsstrahlung amplitude deduced from the Feynman diagrams in Fig.\ref{fig:brem}.
I express it in a relativistic form, but again take  take the non-recoil limit to the amplitude, which gives:
\begin{eqnarray}
\mathcal{M}&\approx&\frac{G_Ve}{\sqrt{2}}\zeta\left(\frac{p\cdot \varepsilon^*}{p\cdot k}-\frac{p_e\cdot \varepsilon^*}{p_e\cdot k}\right)H_\mu L^\mu-\frac{G_Ve}{\sqrt{2}}\left(\frac{\zeta(k^\mu\varepsilon^{*\nu}-k^\nu\varepsilon^{*\mu})+i\epsilon^{\mu\nu\alpha\beta} k_\alpha \varepsilon_\beta^*}{2p_e\cdot k}\right)H_\mu L_\nu\nonumber\\
&\equiv& \mathcal{M}_\text{I}+\mathcal{M}_\text{II}~.
\end{eqnarray}
where 
\begin{equation}
H_\mu=\bar{u}_f(p)\gamma_\mu(1+\lambda\gamma_5)u_i(p)~,~L_{\mu}=\left\{ \begin{array}{ccc}
\bar{u}_{\nu}(p_{\nu})\gamma_{\mu}(1-\gamma_{5})v_{e}(p_{e}) & , & \beta^{+}\\
\bar{u}_{e}(p_{e})\gamma_{\mu}(1-\gamma_{5})v_{\nu}(p_{\nu}) & , & \beta^{-}
\end{array}\right.
\end{equation} 
are the hadronic and leptonic matrix elements of the charged weak current, respectively. Summing over the final spins leads to the following hadronic and leptonic tensors:
\begin{eqnarray}
H_{\mu\nu}&\equiv& \sum_{s_f}H_\mu H_\nu^*=\text{Tr}[(\slashed{p}+m)\gamma_\mu(1+\lambda\gamma_5)\Sigma_i(\slashed{p}+m)\gamma_\nu(1+\lambda\gamma_5)]\nonumber\\
L_{\mu\nu}&\equiv&\sum_{s_e,s_\nu}L_\mu L_\nu^*=8\left\{-g_{\mu\nu}p_e\cdot p_\nu+p_{e\mu}(p_\nu)_\nu+p_{e\nu}(p_\nu)_\mu-i\zeta \epsilon_{\alpha\beta\mu\nu}p_e^\alpha p_\nu^\beta\right\}~,
\end{eqnarray} 
where $\Sigma_i\equiv (1+\gamma_5\slashed{s})/2$ is the parent's spin projection operator, with $s^\mu=(0,\hat{s})$ the parent's spin vector. In what follows, the sum over all final spins is always assumed in the squared amplitude. 

The square of $\mathcal{M}_\text{I}$ needs some extra care:
\begin{equation}
|\mathcal{M}_\text{I}|^2=-\frac{G_V^2e^2}{2}\left(\frac{p}{p\cdot k}-\frac{p_e}{p_e\cdot k}\right)^2H_{\mu\nu}L^{\mu\nu}\equiv |\mathcal{M}_\text{I}|^2_a+|\mathcal{M}_\text{I}|^2_b~,
\end{equation} 
where I split the ``true'' neutrino momentum as $p_\nu=p_\nu'-k$, and $p_\nu'(-k)$ gives the $a(b)$ term respectively; it turns out that the $(a)$ term is the only piece that gives rise to an IR-divergence upon the phase-space integration. So, one can split the full squared amplitude into an IR-divergent piece and a regular piece: $|\mathcal{M}|^2=|\mathcal{M}_\text{I}|^2_a+|\mathcal{M}|^2_\text{reg}$, where
\begin{equation}
|\mathcal{M}|^2_\text{reg}=|\mathcal{M}_\text{I}|^2_b+2\mathfrak{Re}\{\mathcal{M}_\text{I}\mathcal{M}_\text{II}^*\}+|\mathcal{M}_\text{II}|^2~.
\end{equation}
 
\subsection{``Analytic'' part of the outer RC}

Next I plug the squared amplitude into the 4-body decay rate formula in the $\nu'$-formalism, i.e. Eqs.\eqref{eq:Gamma4}, \eqref{eq:phaseapprox}. 
To evaluate the IR-divergent bremsstrahlung contribution I spell out the dot products in $|\mathcal{M}_\text{I}|_a^2$:
\begin{eqnarray}
|\mathcal{M}_\text{I}|_a^2&=&-16G_V^2e^2m^2E_eE_\nu'\left(\frac{p}{p\cdot k}-\frac{p_e}{p_e\cdot k}\right)^2\left\{1+3\lambda^2+(1-\lambda^2)\left(1+\frac{\pnu-E_\nu'}{E_\nu'}\right)\vec{\beta}\cdot\hat{p}_\nu'\right.\nonumber\\
&&\left.+\hat{s}\cdot\left[2\lambda(\lambda\zeta-1)\vec{\beta}-2\lambda(\lambda\zeta+1)\left(1+\frac{\pnu-E_\nu'}{E_\nu'}\right)\hat{p}_\nu'\right]\right\}~,
\end{eqnarray}
where $\hat{p}_\nu'\equiv  \vec{p}_\nu'/\pnu$ is the unit pseudo-neutrino momentum vector. I plug it into Eqs.\eqref{eq:Gamma4}, \eqref{eq:phaseapprox}, and  
the explicit dependence of $p_\nu$, $k$ and $\pnu$ in the expression above allows us to integrate them analytically using the formula in Appendix~\ref{sec:IRdiv}. That gives the following correction to the differential rate :
\begin{eqnarray}
\delta\left(\frac{d\Gamma}{dE_ed\Omega_ed\Omega_\nu'}\right)_{(\text{I}.a)}&=&\frac{G_V^2}{(2\pi)^5}|\vec{p}_e|E_eE_\nu^{\prime 2}\frac{\alpha}{2\pi}\left\{(1+3\lambda^2)\delta_{\text{I}a1}+(1-\lambda^2)(\delta_{\text{I}a1}+\delta_{\text{I}a2})\vec{\beta}\cdot\hat{p}_\nu'\right.\nonumber\\
&&\left.+\hat{s}\cdot\left[2\lambda(\lambda\zeta-1)\delta_{\text{I}a1}\vec{\beta}-2\lambda(\lambda\zeta+1)(\delta_{\text{I}a1}+\delta_{\text{I}a2})\hat{p}_\nu'\right]\right\}~.
\end{eqnarray} 
The function $\delta_{\text{I}a1}$ contains an IR-divergence regularized by a fictitious photon mass $m_\gamma$, which is exactly canceled with that in the virtual outer correction, see Appendix~\ref{sec:virtual}. Combining these two pieces yields the ``analytic'' part of the outer RC:
\begin{eqnarray}
&&\delta\left(\frac{d\Gamma}{dE_ed\Omega_ed\Omega_\nu'}\right)_{v}+\delta\left(\frac{d\Gamma}{dE_ed\Omega_ed\Omega_\nu'}\right)_{(\text{I}.a)}\nonumber\\
&=&\frac{G_V^2}{(2\pi)^5}|\vec{p}_e|E_eE_\nu^{\prime 2}\frac{\alpha}{2\pi}\left\{(1+3\lambda^2)\delta_{\text{an}}+(1-\lambda^2)(\delta_{\text{an}}+\delta_{\text{I}a2}+\delta_{v2})\vec{\beta}\cdot\hat{p}_\nu'\right.\nonumber\\
&&\left.+\hat{s}\cdot\left[2\lambda(\lambda\zeta-1)(\delta_{\text{an}}+\delta_{v2})\vec{\beta}-2\lambda(\lambda\zeta+1)(\delta_{\text{an}}+\delta_{\text{I}a2})\hat{p}_\nu'\right]\right\}~,\label{eq:analytic}
\end{eqnarray} 
where the IR-finite function $\delta_\text{an}$ reads:
\begin{eqnarray}
\delta_\text{an}(E_e,c')&=&\delta_{\text{I}a1}(E_e,c')+\delta_{v1}(E_e)\nonumber\\
&=&-2\left(4-\ln\frac{4E_\nu^{\prime 2}}{m_e^2}\right)\left(\frac{1}{\beta}\tanh^{-1}\beta-1\right)+\frac{3}{2}\ln\frac{m^2}{m_e^2}-\frac{11}{4}+2\ln\left(\frac{1-\beta c'}{1+\beta}\right)\nonumber\\
&&-\frac{1}{\beta}\text{Li}_2\left(\frac{2\beta}{1+\beta}\right)-\frac{1}{\beta}\text{Li}_2\left(\frac{-2\beta}{1-\beta}\right)-\frac{2}{\beta}\text{Li}_2\left(\frac{\beta(c'+1)}{1+\beta}\right)+\frac{2}{\beta}\text{Li}_2\left(\frac{\beta(c'-1)}{1-\beta}\right)\nonumber\\
&&-\frac{2}{\beta}(\tanh^{-1}\beta)^2+2(1+\beta)\tanh^{-1}\beta~,\label{eq:deltaan}
\end{eqnarray}
with $c'\equiv \cos\theta_{e\nu'}=\hat{p}_e\cdot\hat{p}_\nu'$.

\subsection{``Regular'' bremsstrahlung contribution}

Next I evaluate the the contribution from $|\mathcal{M}|_\text{reg}^2$. After applying the replacement \eqref{eq:skreplace} to the quantity $s\cdot k$ in the numerator (see the explanation in Appendix~\ref{sec:spin}), one can organize the squared amplitude as:
\begin{equation}
|\mathcal{M}|_\text{reg}^2\rightarrow 8G_V^2\alpha m^2 E_eE_\nu^{\prime 2}\sum_{ij}\frac{1}{(p\cdot k)^i(p_e\cdot k)^j}\left\{C_{ij}^0+\hat{s}\cdot\left[C_{ij}^\beta \vec{\beta}+C_{ij}^{\nu}\hat{p}_\nu'\right]\right\}~,\label{eq:Mreg2replace}
\end{equation}
where $i,j$ are integers, and the coefficients $C_{ij}^X$ ($X=0,\beta,\nu$) are functions of $E_e,c',\pnu$ and $\lambda$; their explicit expressions can be found in Appendix~\ref{sec:Cij}. In this form, the $p_\nu$- and $k$-integrations can be performed analytically using the formula in Appendix~\ref{sec:Imn}, which give rise to the functions $I_{i,j}$. One is then left with the one-fold integration over $\pnu$, which in principle can also be done analytically with great patience; but for practical purpose it is more efficient to just perform it numerically given that the integration is stable. 
With this, the contribution of $|\mathcal{M}|_\text{reg}^2$ to the differential decay rate in the $\nu'$-formalism reads:
\begin{equation}
\delta\left(\frac{d\Gamma}{dE_ed\Omega_ed\Omega_\nu'}\right)_{\text{reg}}=\frac{G_V^2}{(2\pi)^5}|\vec{p}_e|E_eE_\nu^{\prime 2}\frac{\alpha}{2\pi}\left\{\delta_\text{reg}^0+\hat{s}\cdot\left[\delta_\text{reg}^\beta\vec{\beta}+\delta_\text{reg}^\nu\hat{p}_\nu'\right]\right\}~,\label{eq:reg}
\end{equation} 
where
\begin{equation}
\delta_\text{reg}^X(E_e,c',\lambda)=\sum_{ij}\int_0^{E_\nu'}d\pnu\pnu^2 C_{ij}^X(E_e,c',\pnu,\lambda)I_{i,j}(E_e,c',\pnu)~,~X=0,\beta,\nu~.\label{eq:deltaregX}
\end{equation}

\subsection{Total outer RC in the $\nu'$-formalism}

Adding Eqs.\eqref{eq:analytic} and \eqref{eq:reg} returns the master formula for the total outer RC to the differential decay rate in the $\nu'$-formalism, which is the next central result of this work:
\begin{equation}
\delta\left(\frac{d\Gamma}{dE_ed\Omega_ed\Omega_\nu'}\right)_{\text{outer}}=\frac{G_V^2}{(2\pi)^5}|\vec{p}_e|E_eE_\nu^{\prime 2}\frac{\alpha}{2\pi}\left\{\delta_\text{tot}^0+\hat{s}\cdot\left[\delta_\text{tot}^\beta\vec{\beta}+\delta_\text{tot}^\nu\hat{p}_\nu'\right]\right\}~,\label{eq:total}
\end{equation} 
where
\begin{eqnarray}
\delta_\text{tot}^0(E_e,c',\lambda)&=&(1+3\lambda^2)\delta_\text{an}(E_e,c')+(1-\lambda^2)\left(\delta_\text{an}(E_e,c')+\delta_{\text{I}a2}(E_e)+\delta_{v2}(E_e)\right)\beta c'\nonumber\\
&&+\delta_\text{reg}^0(E_e,c',\lambda)\nonumber\\
\delta_\text{tot}^\beta(E_e,c',\lambda)&=&2\lambda(\lambda\zeta-1)\left(\delta_\text{an}(E_e,c')+\delta_{v2}(E_e)\right)+\delta_\text{reg}^\beta(E_e,c',\lambda)\nonumber\\
\delta_\text{tot}^\nu(E_e,c',\lambda)&=&-2\lambda(\lambda\zeta+1)\left(\delta_\text{an}(E_e,c')+\delta_{\text{I}a2}(E_e)\right)+\delta_\text{reg}^\nu(E_e,c',\lambda)~,
\end{eqnarray}
definitions of the individual functions at the right hand side can be found in Eqs.\eqref{eq:deltaan}, \eqref{eq:deltaregX}, \eqref{eq:deltaIa2} and \eqref{eq:deltav}. This is a fully-analytic formula (apart from a possible numerical integration over $\pnu$ in $\delta_\text{reg}^X$ that can be performed trivially) which is readily applicable to the analysis of neutron and nuclear beta decay data. Advantages of this master formula over other formalisms in literature include the non-necessity to involve  complicated numerical packages, and the explicit cancellation of IR-divergences. 

I stress that, although the results above are based on $J_i=J_f=1/2$, they are equally applicable to $J_i=J_f=0$ by simply taking $\lambda=0$. 

\subsection{Comparison with the $\nu$-formalism}

My master formula should be compared to the result in the traditional $\nu$-formalism~\cite{Sirlin:1967zza,Shann:1971fz,Garcia:1978bq}:
\begin{eqnarray}
\delta\left(\frac{d\Gamma}{dE_ed\Omega_ed\Omega_\nu}\right)_{\text{outer}}&=&\frac{G_V^2}{(2\pi)^5}|\vec{p}_e|E_eE_\nu^{\prime 2}\frac{\alpha}{2\pi}\Bigl\{(1+3\lambda^2)\delta_1+(1-\lambda^2)(\delta_1+\delta_2)\beta c\nonumber\\
&&+\hat{s}\cdot\left[2\lambda(\lambda\zeta-1)(\delta_1+\delta_2)\vec{\beta}-2\lambda(\lambda\zeta+1)\delta_1\hat{p}_\nu\right]\Bigr\}~,\label{eq:outertrad}
\end{eqnarray} 
where $c\equiv \cos\theta_{e\nu}=\hat{p}_e\cdot\hat{p}_\nu$ is to be distinguished from $c'$ in the previous expressions. The functions
\begin{eqnarray}
\delta_1(E_e)&=&3\ln\frac{m}{m_e}-\frac{3}{4}+4\left(\frac{1}{\beta}\tanh^{-1}\beta-1\right)\left(\ln\frac{2E_\nu'}{m_e}+\frac{E_\nu'}{3E_e}-\frac{3}{2}\right)\nonumber\\
&&-\frac{4}{\beta}\text{Li}_2\left(\frac{2\beta}{1+\beta}\right)+\frac{1}{\beta}\tanh^{-1}\beta\left(2+2\beta^2+\frac{E_\nu^{\prime 2}}{6E_e^2}-4\tanh^{-1}\beta\right)\nonumber\\
\delta_2(E_e)&=&2\left(\frac{1-\beta^2}{\beta}\right)\tanh^{-1}\beta+\frac{4E_\nu'(1-\beta^2)}{3\beta^2E_e}\left(\frac{1}{\beta}\tanh^{-1}\beta-1\right)\nonumber\\
&&+\frac{E_\nu^{\prime 2}}{6\beta^2E_e^2}\left(\frac{1-\beta^2}{\beta}\tanh^{-1}\beta-1\right)
\end{eqnarray}
depend only on the electron energy and not the angles. Comparing Eqs.\eqref{eq:total} and \eqref{eq:outertrad}, the former utilizes a measurable $c'=\cos\theta_{e\nu'}=\hat{p}_e\cdot\hat{p}_\nu'$ while the latter depends on an unmeasured quantity $c=\cos\theta_{e\nu}=\hat{p}_e\cdot\hat{p}_\nu$. 
An important consistency check is that the two expressions should give rise to the same result for $\delta(d\Gamma/dE_ed\Omega_e)_\text{outer}$; in other words, if the recoiled daughter nucleus is not detected in the experiment (e.g. during the measurement of the beta spectrum or the beta asymmetry parameter $A$ that depend only on $\vec{p}_e$), then the $\nu$- and $\nu'$-formalisms make no difference. This implies the following relations:
\begin{eqnarray}
\frac{1}{2}\int_{-1}^{+1}dc'\delta_\text{tot}^0(E_e,c',\lambda)&=&(1+3\lambda^2)\delta_1(E_e)\nonumber\\
\frac{1}{2}\int_{-1}^{+1}dc'\left[\delta_\text{tot}^\beta(E_e,c',\lambda)+\frac{c'}{\beta}\delta_\text{tot}^\nu(E_e,c',\lambda)\right]&=&2\lambda(\lambda\zeta-1)\left(\delta_1(E_e)+\delta_2(E_e)\right)~.
\end{eqnarray}
One can check numerically that the two relations above are \textit{exactly} satisfied, which is a solid proof of the correctness of all the analytic formula presented in this paper.

\begin{figure}
	\begin{centering}
		\includegraphics[scale=0.47]{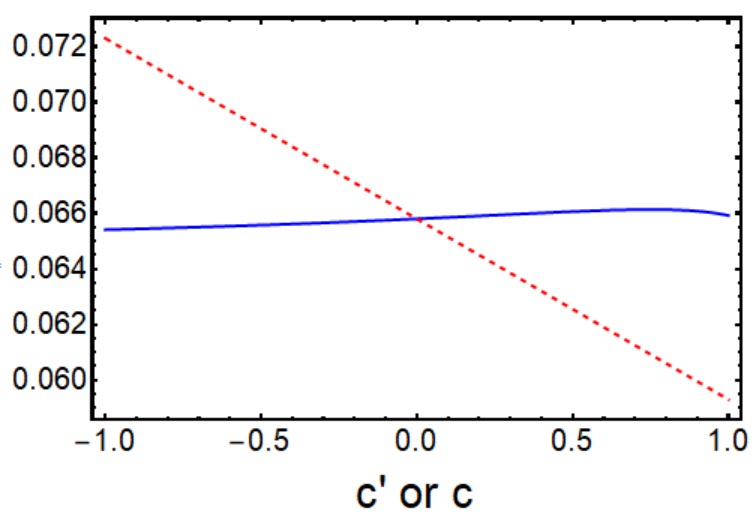}
		\includegraphics[scale=0.45]{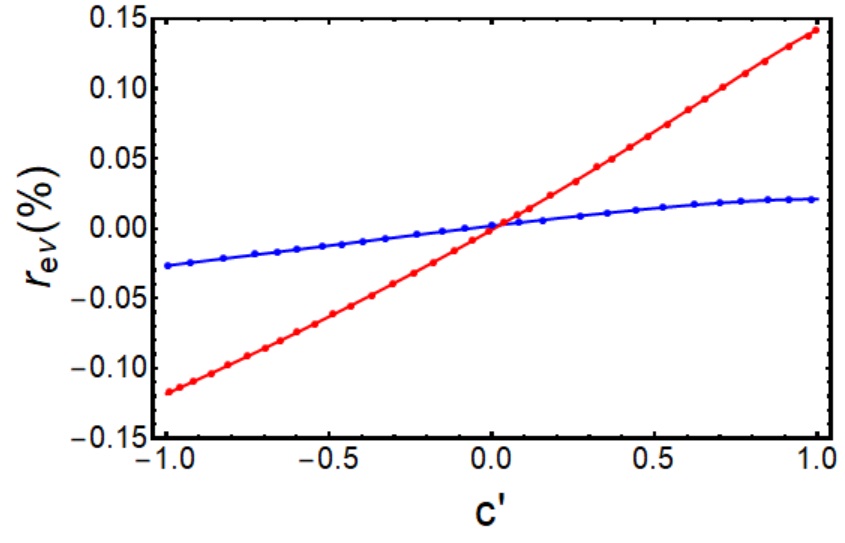}
		\par\end{centering}
	\caption{\label{fig:compare}Left: $(\alpha/2\pi)\delta_\text{tot}^0$ (blue, solid) as a function of $c'$ versus $(\alpha/2\pi)[(1+3\lambda^2)\delta_1+(1-\lambda^2)(\delta_1+\delta_2)\beta c]$ (red, dashed) as a function of $c$ at $E_e=0.8E_m$. Right: The function $r_{e\nu}$ at $E_e=m_e+0.2(E_m-m_e)$ (blue) and $E_e=m_e+0.8(E_m-m_e)$ (red). Solid line: My result; dots: Discrete points extracted from Fig.8 in Ref.\cite{Gluck:2022ogz}. Here I take $\lambda=-1.27$.  }
\end{figure}

\begin{table}
	\begin{centering}
		\begin{tabular}{|c|cccccccc|}
			\hline 
			\multicolumn{1}{|c}{$c'$} &  &  &  & $r_{e\nu}$ & (\%) &  &  & \tabularnewline
			\hline 
			\hline 
			0.9 & 0.021 & 0.038 & 0.056 & 0.075 & 0.093 & 0.112 & 0.129 & 0.146\tabularnewline
			\cline{1-1} 
			0.7 & 0.018 & 0.032 & 0.046 & 0.060 & 0.074 & 0.087 & 0.100 & 0.112\tabularnewline
			\cline{1-1} 
			0.5 & 0.014 & 0.024 & 0.034 & 0.044 & 0.053 & 0.061 & 0.069 & 0.077\tabularnewline
			\cline{1-1} 
			0.3 & 0.010 & 0.016 & 0.021 & 0.026 & 0.031 & 0.036 & 0.040 & 0.044\tabularnewline
			\cline{1-1} 
			0.1 & 0.005 & 0.006 & 0.008 & 0.009 & 0.011 & 0.012 & 0.012 & 0.013\tabularnewline
			\cline{1-1} 
			-0.1 & -0.001 & -0.003 & -0.005 & -0.007 & -0.009 & -0.012 & -0.014 & -0.016\tabularnewline
			\cline{1-1} 
			-0.3 & -0.007 & -0.012 & -0.018 & -0.023 & -0.029 & -0.034 & -0.039 & -0.044\tabularnewline
			\cline{1-1} 
			-0.5 & -0.012 & -0.021 & -0.030 & -0.039 & -0.047 & -0.055 & -0.063 & -0.071\tabularnewline
			\cline{1-1} 
			-0.7 & -0.018 & -0.030 & -0.042 & -0.054 & -0.065 & -0.076 & -0.086 & -0.096\tabularnewline
			\cline{1-1} 
			-0.9 & -0.024 & -0.039 & -0.054 & -0.068 & -0.082 & -0.095 & -0.108 & -0.120\tabularnewline
			\hline 
			$x_{e}$ & \multicolumn{1}{c|}{0.2} & \multicolumn{1}{c|}{0.3} & \multicolumn{1}{c|}{0.4} & \multicolumn{1}{c|}{0.5} & \multicolumn{1}{c|}{0.6} & \multicolumn{1}{c|}{0.7} & \multicolumn{1}{c|}{0.8} & 0.9\tabularnewline
			\hline 
		\end{tabular}
		\par\end{centering}
	\caption{\label{tab:renu}My result of $r_{e\nu}(E_e,c')$ with $E_e=m_e+x_e(E_m-m_e)$, which should be compared to Table~V in Ref.\cite{Gluck:1992tg}.}
	
\end{table}

I argued before that, a mis-identification of $\hat{p}_\nu$ by $\hat{p}_\nu'$ in the $\nu$-formalism would lead to an error of the order $\alpha/\pi\sim 10^{-3}$. To see this, let us compare the numerical results deduced from Eqs.\eqref{eq:total} and \eqref{eq:outertrad} for the case of neutron decay: $\zeta=-1$, $m_i=m_n$, $m_f=m_p$. 
I start from the spin-independent part of the outer RC, namely $(\alpha/2\pi)\delta_\text{tot}^0$ in the $\nu'$-formalism and $(\alpha/2\pi)[(1+3\lambda^2)\delta_1+(1-\lambda^2)(\delta_1+\delta_2)\beta c]$ in the $\nu$-formalism; they affect the precise extraction of the neutrino-electron correlation coefficient $a$. From the first diagram in Fig.\ref{fig:compare} one sees that the dependence of these functions on their respective cosines indeed shows a difference at the order $10^{-3}$. Also, to compare with existing literature, in the second diagram I plot the the $\{E_e,c'\}$ Dalitz distribution of the outer RC:
\begin{equation}
r_{e\nu}(E_e,c')\equiv\frac{\delta(d\Gamma/dE_edc')_\text{outer}}{(d\Gamma/dE_edc')_\text{tree}}-\frac{\delta(d\Gamma/dE_e)_\text{outer}}{(d\Gamma/dE_e)_\text{tree}}\approx \frac{\alpha}{2\pi}\left[\frac{\delta_\text{tot}^0(E_e,c',\lambda)}{1+3\lambda^2+(1-\lambda^2)\beta c'}-\delta_1(E_e)\right]
\end{equation}
at two different $E_e$; one finds excellent agreement with Fig.8 of Ref.\cite{Gluck:2022ogz}. One also notices that a table of $r_{e\nu}(E_e,c')$ was given in an earlier paper~\cite{Gluck:1992tg}, but using my analytic expressions one finds that some entries in that table are off by an absolute value of $\sim 0.01\%$. My new results are displayed in Table~\ref{tab:renu}, but for practical applications one should of course recompute everything using my given formula instead of just applying the table. 

\begin{figure}
	\begin{centering}
		\includegraphics[scale=0.7]{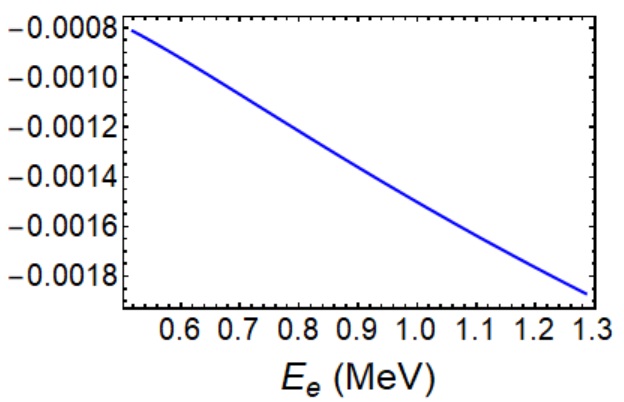}
		\par\end{centering}
	\caption{\label{fig:delta_nu}
	The difference $(\alpha/2\pi)[\bar{\delta}_\nu'(E_e,\lambda)-2\lambda(\lambda-1)\delta_1(E_e)]$ as a function of $E_e$. Again I take $\lambda=-1.27$.  }
\end{figure}

Next I study a more interesting quantity, namely the outer RC needed to extract the neutrino asymmetry parameter $B$. At $\mathcal{O}(\alpha^0)$ (which does not distinguish $\nu'$ from $\nu$), a possible way to extract $B$ is to flip the neutron spin and integrate over $\Omega_e$:
\begin{equation}
\left.\left(\frac{d\Gamma}{dE_ed\Omega_\nu^{(\prime)}}\right)_\text{tree}\right|_{\hat{s}}-\left.(\dots)\right|_{-\hat{s}}\propto B\hat{s}\cdot\hat{p}_\nu^{(\prime)}~.
\end{equation}
However, outer correction to the expression above takes very different forms in the $\nu$- and $\nu'$-formalism:
\begin{eqnarray}
\delta\left.\left(\frac{d\Gamma}{dE_ed\Omega_\nu}\right)_\text{outer}\right|_{\hat{s}}-\left.(\dots)\right|_{-\hat{s}}&\propto& \frac{\alpha}{2\pi}\times 2\lambda(\lambda-1)\delta_1(E_e)\hat{s}\cdot\hat{p}_\nu\nonumber\\
\delta\left.\left(\frac{d\Gamma}{dE_ed\Omega_\nu'}\right)_\text{outer}\right|_{\hat{s}}-\left.(\dots)\right|_{-\hat{s}}&\propto&\frac{\alpha}{2\pi}\times\frac{1}{2}\int_{-1}^{+1}dc'\left[\delta_\text{tot}^\nu(E_e,c',\lambda)+\beta c'\delta_\text{tot}^\beta(E_e,c',\lambda)\right]\hat{s}\cdot\hat{p}_\nu'\nonumber\\
&\equiv&\frac{\alpha}{2\pi}\bar{\delta}_\nu'(E_e,\lambda)\hat{s}\cdot\hat{p}_\nu'
\end{eqnarray}
One sees that, unlike the $\nu$-formalism that probes only the $\hat{s}\cdot \hat{p}_\nu$ term in the outer correction, the $\nu'$-formalism probes both the $\hat{s}\cdot\hat{p}_\nu'$ and $\hat{s}\cdot\vec{\beta}$ terms, because both their coefficients are functions of $c'$. In Fig.\ref{fig:delta_nu} I plot the difference between the function $(\alpha/2\pi)\bar{\delta}_\nu'(E_e,\lambda)$ (defined above) and $(\alpha/2\pi)\times2\lambda(\lambda-1)\delta_1(E_e)$, and see that it is again of the order $10^{-3}$. This means, a mis-identification of $\hat{p}_\nu$ by $\hat{p}_\nu'$ in the $\nu$-formalism would lead to a $10^{-3}$ error in the extraction of $B$.

\section{Radiative corrections in the recoil formalism}

In some experiments such as aCORN and Nab~\cite{Fry:2018kvq}, it is $E_f$ instead of $\Omega_\nu'$ that is being measured; in this case the recoil formalism, and not $\nu'$, is the more appropriate description for the outer RC. So, for the sake of completeness I provide also the set of analytic expressions to compute the outer RC of the differential decay rate in the recoil formalism, in full analogy to those in the $\nu'$-formalism from the previous sections.  

First, it is obvious that one has no access to spin-dependent correlations if only $E_e$ and $E_f$ are measured, because there would be no vector to contract with $\hat{s}$ as all solid angles of momenta are integrated out. Thus, for decays of polarized nuclei it is more natural to adopt the $\nu'$-formalism where the solid angles $\Omega_e$ and $\Omega_\nu'$ are fixed, whereas in the recoil formalism I shall restrict myself to \textit{unpolarized} nuclei. I start with the tree-level squared amplitude in the non-recoil limit, Eq.\eqref{eq:Mtree2}, and recast the variables in terms of 4-vector products:
\begin{equation}
E_e=\frac{p_i\cdot p_e}{m_i}~,~E_\nu'=\frac{p_i\cdot p_\nu'}{m_i}~,~\vec{p}_e\cdot\vec{p}_\nu'=\frac{p_i\cdot p_e p_i\cdot p_\nu'}{m_i^2}-p_e\cdot p_\nu'~,\label{eq:todots}
\end{equation}
which allows to express the squared amplitude as a function of $\{y,z\}$ (recall that $x=0$ for 3-body decay). That gives the tree-level differential rate:
\begin{equation}
\left(\frac{d\Gamma}{dydz}\right)_\text{tree}^{\mathcal{D}_3}\approx \frac{G_V^2m^5}{(4\pi)^3}\left\{(1+3\lambda^2)y(2-y-z)+(1-\lambda^2)\left[y(2-y-z)+2(r_e-r_f+z-1)\right]\right\}~,
\end{equation}
where the superscript $\mathcal{D}_3$ denotes that the expression survives only if $\{y,z\}\in\mathcal{D}_3$. 

The determination of the outer RC proceeds in the exact same way as before: For the virtual corrections, I take again the results in Refs.\cite{Sirlin:1967zza,Shann:1971fz,Garcia:1978bq} and use Eq.\eqref{eq:todots} to recast them in terms of $\{y,z\}$; this contribution survives only in the $\mathcal{D}_3$ region. For the bremsstrahlung contribution, I follow Sec.\ref{sec:Brsq} and split the squared amplitude into two pieces: 
\begin{equation}
|\mathcal{M}|^2=|\mathcal{M}_\text{I}|_a^2+|\mathcal{M}|_\text{reg}^2~,
\end{equation} 
and plug it into Eq.\eqref{eq:phaseexact};
the $x$-integral is evaluated analytically for $|\mathcal{M}_\text{I}|^2_a$ and numerically for $|\mathcal{M}|^2_\text{reg}$. The IR-divergence occurs in the $x$-integral of $|\mathcal{M}_\text{I}|_a^2$ in the $\mathcal{D}_3$ region, which analytic result can be found in Appendix D of Ref.\cite{Seng:2021wcf}; it combines with the virtual correction to yield an IR-finite outcome. Meanwhile, for the ``regular'' piece I expand the squared amplitude as:
\begin{equation}
|\mathcal{M}|_\text{reg}^2=4G_V^2\alpha m^2\sum_{ij}\frac{1}{(p\cdot k)^i(p_e\cdot k)^j}d_{ij}~,\label{eq:recoilexpand}
\end{equation} 
where the expansion coefficients are just $d_{ij}=2E_e E_\nu^{\prime 2} C_{ij}^0$; nevertheless, I still provide their analytic expressions in terms of $\{x,y,z\}$ in Appendix~\ref{sec:dij} for the convenience of readers. The $p_\nu$- and $k$-integrals can be performed analytically using Appendix~\ref{sec:Imn}.

Combining all the above, I obtain the total outer RC in the $\mathcal{D}_3$ and $\mathcal{D}_{4-3}$ region as follows:
\begin{eqnarray}
\delta\left(\frac{d\Gamma}{dydz}\right)_\text{outer}^{\mathcal{D}_3}&=&\frac{G_V^2m^5}{(4\pi)^3}\frac{\alpha}{2\pi}\left\{(1+3\lambda^2)y(2-y-z)\delta_\text{an}^r+2(1-\lambda^2)\alpha_+\delta_{\text{Ia2}}^r\right.\nonumber\\
&&\left.+(1-\lambda^2)\left[y(2-y-z)+2(r_e-r_f+z-1)\right](\delta_\text{an}^r+\delta_{v2})+\delta_\text{reg}^{r,\mathcal{D}_3}\right\}\nonumber\\
\delta\left(\frac{d\Gamma}{dydz}\right)_\text{outer}^{\mathcal{D}_{4-3}}&=&\frac{G_V^2m^5}{(4\pi)^3}\frac{\alpha}{2\pi}\left\{\delta_{\text{Ia2}}^r\ln\frac{\alpha_+}{\alpha_-}\left[(1+3\lambda^2)y(2-y-z)+(1-\lambda^2)\left(y(2-y-z)\right.\right.\right.\nonumber\\
&&\left.\left.\left.+2(r_e-r_f+z-1)\right)\right]+2(1-\lambda^2)(\alpha_+-\alpha_-)\delta_{\text{Ia2}}^r+\delta_\text{reg}^{r,\mathcal{D}_{4-3}}\right\}~,\label{eq:raterecoil}
\end{eqnarray}
where the functions at the right hand side are (the extra superscript $r$ stands for ``recoil''):
\begin{eqnarray}
\delta_\text{an}^r(y,z)&=&\left(\frac{2}{\beta}\tanh^{-1}\beta+\frac{1}{2}\right)\ln\frac{m^2}{m_e^2}-\frac{11}{4}+\left(\frac{2}{\beta}\tanh^{-1}\beta-1\right)\ln\frac{m^2\alpha_+^2}{4P_0^2}\nonumber\\
&&+\ln\frac{(r_e-r_f+z-1)^2}{\alpha_+^2}-\frac{1}{\beta}\text{Li}_2\left(\frac{2\beta}{1+\beta}\right)-\frac{1}{\beta}\text{Li}_2\left(-\frac{2\beta}{1-\beta}\right)\nonumber\\
&&-\frac{2}{\beta}\text{Li}_2\left(\frac{\beta}{1+\beta}\left(\frac{P_1}{P_0}+1\right)\right)+\frac{2}{\beta}\text{Li}_2\left(\frac{\beta}{1-\beta}\left(\frac{P_1}{P_0}-1\right)\right)\nonumber\\
&&-\frac{2}{\beta}\left(\tanh^{-1}\beta\right)^2+2\beta\tanh^{-1}\beta\nonumber\\
\delta_{\text{Ia2}}^r(y)&=&4\left(\frac{1}{\beta}\tanh^{-1}\beta-1\right)\nonumber\\
\delta_{v2}(y)&=&\frac{2(1-\beta^2)}{\beta}\tanh^{-1}\beta\nonumber\\
\delta_\text{reg}^{r,\mathcal{D}_3}(y,z)&=&\sum_{ij}\int_0^{\alpha_+}dx d_{ij}I_{i,j}\nonumber\\
\delta_\text{reg}^{r,\mathcal{D}_{4-3}}(y,z)&=&\sum_{ij}\int_{\alpha_-}^{\alpha_+}dx d_{ij}I_{i,j}~,
\end{eqnarray}
with
\begin{equation}
P_0=\frac{m}{2}(2-y-z)~,~P_1=\frac{m}{2\beta y}\left[y(2-y-z)+2(r_e-r_f+z-1)\right]~,
\end{equation}
and the electron speed $\beta=\sqrt{y^2-4r_e}/y$.

\begin{figure}
	\begin{centering}
		\includegraphics[scale=0.5]{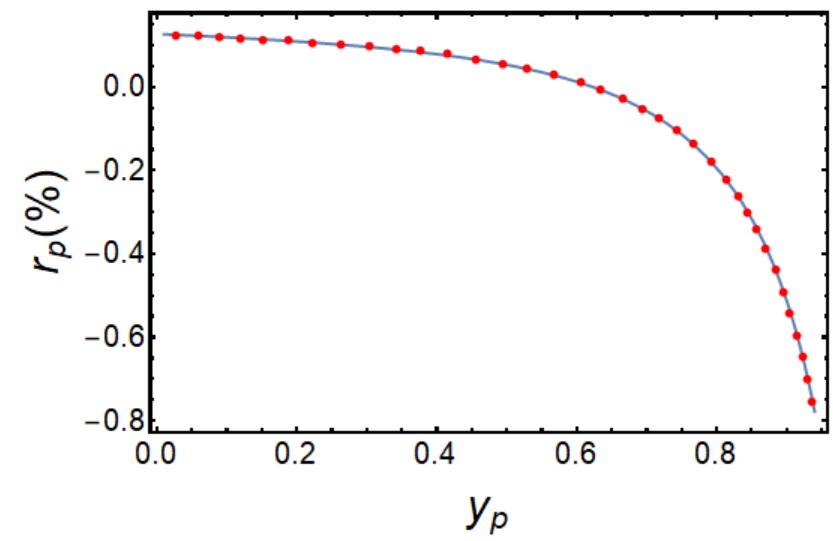}
		\includegraphics[scale=0.6]{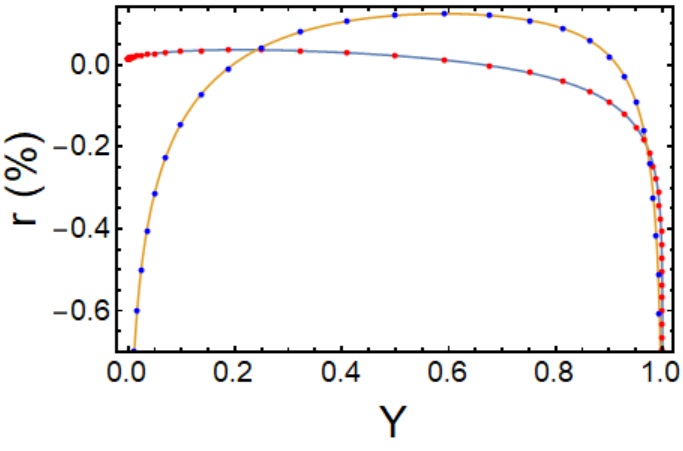}
		\par\end{centering}
	\caption{\label{fig:rp}Left: The function $r_p$ with respect to the proton kinetic energy $T=y_p T_\text{max}=y_p((m_n-m_p)^2-m_e^2)/(2m_n)$, with $\lambda=-1.27$; solid line is my result and dots are discrete points extracted from Fig.7 in Ref.\cite{Gluck:2022ogz}. Right: My predictions of the function $r$ at $X=0.2$ (blue solid line) and $X=0.8$ (yellow solid line); dots are corresponding discrete points extracted from Fig.6(a) in Ref.\cite{Gluck:2022ogz}. }
\end{figure}

Eq.\eqref{eq:raterecoil} is the third central result of this work; a simple consistency check is that it must reproduce the same outer RC to the beta spectrum as in Sirlin's approach, up to small recoil corrections (on top of the RC). This implies:
\begin{equation}
\frac{\delta\left(d\Gamma/dy\right)_\text{outer}}{\left(d\Gamma/dy\right)_\text{tree}}=\frac{\int dz\delta\left(d\Gamma/dydz\right)_\text{outer}}{\int dz\left(d\Gamma/dydz\right)_\text{tree}}\approx \frac{\alpha}{2\pi}\delta_1(E_e)~,~2\sqrt{r_e}<y<1+r_e-r_f~.
\end{equation}
Substituting the neutron decay parameters, one finds that this relation is satisfied to a $<0.1\%$ relative error, which proves the correctness of all the analytic formula. To provide further cross-check, I compute two more functions: The first is the relative RC $r_p$ to the proton spectrum in neutron decay defined in Ref.\cite{Gluck:2022ogz}, which reads:
\begin{equation}
r_p(T)\equiv\frac{\delta\left(d\Gamma/dz\right)_\text{outer}}{\left(d\Gamma/dz\right)_\text{tree}}-\frac{\delta\Gamma_\text{outer}}{\Gamma_\text{tree}}~,
\end{equation} 
where I use the proton kinetic energy $T=m_n z/2-m_p$ as the variable. The second is the relative RC of the Dalitz distribution $\{y,z\}$ in the $\mathcal{D}_3$ region as defined in the same reference:
\begin{equation}
r(y,z)\equiv \frac{\delta(d\Gamma/dydz)_\text{outer}^{\mathcal{D}_3}}{(d\Gamma/dydz)_\text{tree}}-\frac{\delta(d\Gamma/dy)_\text{outer}^{\mathcal{D}_3}}{(d\Gamma/dy)_\text{tree}}~,
\end{equation}
where the results are expressed in terms of two new variables $\{X,Y\}$ (which correspond to $\{x,y\}$ in Ref.\cite{Gluck:2022ogz}), defined through:
\begin{equation}
y=\frac{2}{m_n}(m_e+(E_m-m_e)X)~,~z=z_\text{min}+(z_\text{max}-z_\text{min})Y~,
\end{equation}
with $z_\text{min}=a(y)-b(y)$, $z_\text{max}=a(y)+b(y)$ in the $\mathcal{D}_3$ region. The outcomes are plotted in Fig.\ref{fig:rp} and one again observes excellent agreement with the results in Ref.\cite{Gluck:2022ogz} obtained using numerical packages.

\section{Summary}

In this paper I carefully analyze the theory structure of the so-called $\nu'$-formalism, which is designed to bypass the conceptual problem in the traditional treatment of outer RC by Sirlin, Shann and Garcia-Maya. The formalism itself has already appeared in many literature and is not at all a new concept, but this paper provides the most elegant representation of the 4-body phase space formula consistent to the non-recoil approximation. With this master formula, I derive the full set of analytic expressions needed for the evaluation of the outer RC to the differential rate of a $J_i=J_f=1/2$ (or 0) nuclear $\beta^\pm$ decay. The results are explicitly $m_\gamma$-independent, and require only a minimal amount of numerical integration. I compare the outcome with existing literature and discuss its implications on $\vec{p}_\nu'$-dependent observables. Similar analytic expressions in the recoil formalism are also derived assuming unpolarized nuclei. 

While the master formula for the 4-body phase space is completely general, I restrict myself to spin-half systems which allows us to make use of the elegant spinor technique in field theory that simplifies the squared amplitude. Generalization to generic nuclear beta decays with arbitrary $J_{i,f}$ may involve, apart from more complicated analytic expressions for existing terms, new spin-dependent correlation structures such as the $c$-coefficient~\cite{Jackson:1957zz}. Its outer RC may be studied either using traditional approaches focusing on the ``convention term'' contribution~\cite{Meister:1963zz}, or using the more recent effective field theory description~\cite{Cirigliano:2023fnz,Hill:2023acw,Hill:2023bfh}. This will be worked out in a follow-up work.

\begin{acknowledgments}
	
I thank Ferenc Gl\"{u}ck for many useful discussions and for cross-checking my results.
I was supported in
part by the U.S. Department of Energy (DOE), Office of Science, Office of Nuclear Physics, under the FRIB Theory Alliance award DE-SC0013617, and by the DOE grant DE-FG02-97ER41014. I acknowledge support from the DOE Topical Collaboration ``Nuclear Theory for New Physics'', award No. DE-SC0023663. 
  
\end{acknowledgments}

\begin{appendix}
	
\section{\label{sec:IRdiv}IR-divergent phase space integral}

In the $\nu'$-formalism, the only IR-divergent phase space integral reads:
\begin{equation}
\int_0^{E_\nu'}d\pnu\pnu^2\int\frac{d^3k}{(2\pi)^32E_k}\frac{d^3p_\nu}{(2\pi)^32E_\nu}(2\pi)^4\delta^{(4)}(p_\nu'-p_\nu-k)\left(\frac{p}{p\cdot k}-\frac{p_e}{p_e\cdot k}\right)^2\equiv-\frac{E_\nu'}{8\pi}\delta_{\text{I}a1}~.\label{eq:IRdiv}
\end{equation}
The IR-divergence arises from the $\pnu$-integral at $\pnu\rightarrow E_\nu'$. One may evaluate this integral using the method outlined in Ref.\cite{Seng:2021wcf}, namely to apply dimensional regularization of the photon momentum~\cite{Marciano:1974tv}:
\begin{equation}
\frac{d^3k}{(2\pi)^32E_k}\rightarrow \mu^{4-d}\frac{d^{d-1}k}{(2\pi)^{d-1}2E_k}~.
\end{equation}
A important advantage of this method is that the massless on-shell condition $k^2=0$ still holds, which greatly simplifies the intermediate steps. After finishing the evaluation, one may choose to switch back to the photon-mass prescription by the following matching:
\begin{equation}
\frac{2}{4-d}-\gamma_E+\ln 4\pi\rightarrow \ln\frac{m_\gamma^2}{\mu^2}~.
\end{equation}
With this, I obtain the result of the integral as:
\begin{eqnarray}
\delta_{\text{I}a1}(E_e,c')&=&-2\left(4-\ln\frac{4E_\nu^{\prime 2}}{m_\gamma^2}\right)\left(\frac{1}{\beta}\tanh^{-1}\beta-1\right)+2\tanh^{-1}\beta+2\ln\left(\frac{1-\beta c'}{1+\beta}\right)\nonumber\\
&&+\frac{1}{\beta}\text{Li}_2\left(\frac{2\beta}{1+\beta}\right)-\frac{1}{\beta}\text{Li}_2\left(\frac{-2\beta}{1-\beta}\right)-\frac{2}{\beta}\text{Li}_2\left(\frac{\beta(c'+1)}{1+\beta}\right)\nonumber\\
&&+\frac{2}{\beta}\text{Li}_2\left(\frac{\beta(c'-1)}{1-\beta}\right)~.
\end{eqnarray}
 
For terms proportional to $\hat{p}_\nu'$ in $|\mathcal{M}_\text{I}|^2_a$, the involved integral is Eq.\eqref{eq:IRdiv} but with an extra factor $1+(\pnu-E_\nu')/E_\nu'$ multiplied to the integrand. The integration returns $-(E_\nu'/8\pi)(\delta_{\text{I}a1}+\delta_{\text{I}a2})$, where
\begin{equation}
\delta_{\text{I}a2}(E_e)=4(1-\ln 4)\left(\frac{1}{\beta}\tanh^{-1}\beta-1\right)~.\label{eq:deltaIa2}
\end{equation}

\section{\label{sec:virtual}Virtual outer corrections}

\begin{figure}
	\begin{centering}
		\includegraphics[scale=0.8]{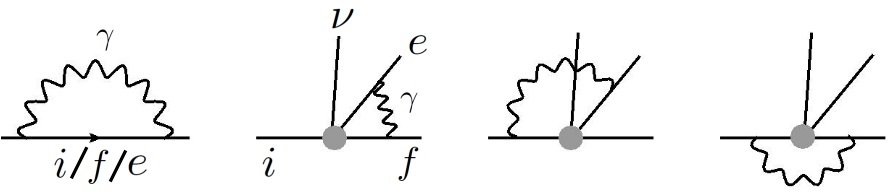}
		\par\end{centering}
	\caption{\label{fig:loop} Self-energy and one-particle irreducible diagrams that contribute to the outer RC.  }
\end{figure}

The SM one-loop RC to beta decay contribute to  three quantities: (1) The Fermi function~\cite{Fermi:1934hr} that describes the Coulomb interaction between the outgoing electron and the daughter nucleus; (2) The ``outer'' RC that summarizes all the non-Fermi and $E_e$-dependent corrections; (3) The ``inner'' RC which only effect is to renormalize the effective weak coupling constants. In this paper I am only interested in the outer RC, which is obtained by computing the Feynman diagrams in Fig.\ref{fig:loop}, taking the nuclei as point-like particles. For a $J_i=J_f=1/2$ decay with polarized parent nucleus, the outcome in the non-recoil limit reads~\cite{Sirlin:1967zza,Shann:1971fz,Garcia:1978bq}:
\begin{eqnarray}
\delta\left(\frac{d\Gamma}{dE_ed\Omega_ed\Omega_\nu'}\right)_v&=&\frac{G_V^2}{(2\pi)^5}|\vec{p}_e|E_eE_\nu^{\prime 2}\frac{\alpha}{2\pi}\left\{(1+3\lambda^2)\delta_{v1}+(1-\lambda^2)(\delta_{v1}+\delta_{v2})\vec{\beta}\cdot\hat{p}_\nu'\right.\nonumber\\
&&\left.+\hat{s}\cdot\left[2\lambda(\lambda\zeta-1)(\delta_{v1}+\delta_{v2})\vec{\beta}-2\lambda(\lambda\zeta+1)\delta_{v1}\hat{p}_\nu'\right]\right\}~,
\end{eqnarray}
where
\begin{eqnarray}
\delta_{v1}(E_e)&=&-2\ln\frac{m_e^2}{m_\gamma^2}\left(\frac{1}{\beta}\tanh^{-1}\beta-1\right)+\frac{3}{2}\ln\frac{m^2}{m_e^2}-\frac{11}{4}-\frac{2}{\beta}\text{Li}_2\left(\frac{2\beta}{1+\beta}\right)\nonumber\\
&&-\frac{2}{\beta}(\tanh^{-1}\beta)^2+2\beta\tanh^{-1}\beta\nonumber\\
\delta_{v2}(E_e)&=&\frac{2(1-\beta^2)}{\beta}\tanh^{-1}\beta~.\label{eq:deltav}
\end{eqnarray}

\section{\label{sec:Imn}Analytic formula for $p_\nu$- and $k$-integration}

If one drops the parent's spin vector $s^\mu$, then there are only three vectors that can be dotted to the photon momentum $k$ in the squared amplitude: $p$, $p_e$ and $p_\nu'$; but momentum conservation $p_\nu'-k=p_\nu$ implies $2p_\nu'\cdot k=p_\nu^{\prime 2}$ assuming massless neutrino. 
The $p_\nu$- and $k$-integrations in $|\mathcal{M}|^2_\text{reg}$ can be represented in terms of the functions below:
\begin{equation}
I_{i,j}(E_e,c',\pnu)\equiv \int\frac{d^3k}{(2\pi)^32E_k}\frac{d^3p_\nu}{(2\pi)^32E_\nu}(2\pi)^4\delta^{(4)}(p_\nu'-p_\nu-k)\frac{1}{(p\cdot k)^i(p_e\cdot k)^j}~,
\end{equation}
where $i,j$ are integers. Analytic expressions of the functions needed in this work can be found in Ref.\cite{Ginsberg:1969jh} (but beware of the difference in the overall normalization):
\begin{eqnarray}
I_{0,0}&=&\frac{1}{8\pi}\nonumber\\
I_{-1,0}&=&\frac{\alpha_p}{16\pi}\nonumber\\
I_{1,0}&=&\frac{1}{8\pi\beta_p}\ln\frac{\alpha_p+\beta_p}{\alpha_p-\beta_p}\nonumber\\
I_{0,1}&=&\frac{1}{8\pi\beta_e}\ln\frac{\alpha_e+\beta_e}{\alpha_e-\beta_e}\nonumber\\
I_{2,0}&=&\frac{1}{2\pi m^2p_\nu^{\prime 2}}\nonumber\\
I_{0,2}&=&\frac{1}{2\pi m_e^2p_\nu^{\prime 2}}\nonumber\\
I_{1,1}&=&\frac{1}{4\pi\gamma_{pe}p_\nu^{\prime 2}}\ln\frac{p\cdot p_e+\gamma_{pe}}{p\cdot p_e-\gamma_{pe}}\nonumber\\
I_{1,-1}&=&\frac{1}{8\pi}\left(\frac{(pp_e:p_\nu')}{\beta_p^2}+\frac{p_\nu^{\prime 2}(p_ep_\nu':p)}{2\beta_p^3}\ln\frac{\alpha_p+\beta_p}{\alpha_p-\beta_p}\right)\nonumber\\
I_{-1,1}&=&\frac{1}{8\pi}\left(\frac{(p_ep:p_\nu')}{\beta_e^2}+\frac{p_\nu^{\prime 2}(pp_\nu':p_e)}{2\beta_e^3}\ln\frac{\alpha_e+\beta_e}{\alpha_e-\beta_e}\right)\nonumber\\
I_{2,-1}&=&\frac{1}{8\pi}\left(\frac{2(p_ep_\nu':p)}{m^2\beta_p^2}+\frac{(pp_e:p_\nu')}{\beta_p^3}\ln\frac{\alpha_p+\beta_p}{\alpha_p-\beta_p}\right)\nonumber\\
I_{-1,2}&=&\frac{1}{8\pi}\left(\frac{2(pp_\nu':p_e)}{m_e^2\beta_e^2}+\frac{(p_ep:p_\nu')}{\beta_e^3}\ln\frac{\alpha_e+\beta_e}{\alpha_e-\beta_e}\right)\nonumber\\
I_{-2,1}&=&\frac{1}{8\pi}\left[\frac{(p_\nu^{\prime 2})^2(pp_\nu':p_e)^2}{4\beta_e^5}\ln\frac{\alpha_e+\beta_e}{\alpha_e-\beta_e}+\frac{p_\nu^{\prime 2}(p_ep:p_\nu')(pp_\nu':p_e)}{\beta_e^4}+\frac{\alpha_e(p_ep:p_\nu')^2}{2\beta_e^4}\right.\nonumber\\
&&\left.+\left(\frac{\beta_e^2\beta_p^2-(p_ep:p_\nu')^2}{4\beta_e^4}\right)\left(\alpha_e-\frac{m_e^2p_\nu^{\prime 2}}{2\beta_e}\ln\frac{\alpha_e+\beta_e}{\alpha_e-\beta_e}\right)\right]\nonumber\\
I_{-2,2}&=&\frac{1}{8\pi}\left[\frac{p_\nu^{\prime 2}(pp_\nu':p_e)^2}{m_e^2\beta_e^4}+\frac{p_\nu^{\prime 2}(pp_\nu':p_e)(p_ep:p_\nu')}{\beta_e^5}\ln\frac{\alpha_e+\beta_e}{\alpha_e-\beta_e}+\frac{(p_ep:p_\nu')^2}{\beta_e^4}\right.\nonumber\\
&&\left.-\left(\frac{\beta_e^2\beta_p^2-(p_ep:p_\nu')^2}{2\beta_e^4}\right)\left(2-\frac{\alpha_e}{\beta_e}\ln\frac{\alpha_e+\beta_e}{\alpha_e-\beta_e}\right)\right]~.
\end{eqnarray}	
Here I have defined:
\begin{eqnarray}
&&\alpha_p\equiv p\cdot p_\nu'~,~\alpha_e\equiv p_e\cdot p_\nu'~,~\beta_p\equiv\sqrt{\alpha_p^2-m^2p_\nu^{\prime 2}}~,~\beta_e\equiv\sqrt{\alpha_e^2-m_e^2p_\nu^{\prime 2}}~,\nonumber\\
&&\gamma_{pe}\equiv \sqrt{(p\cdot p_e)^2-m^2m_e^2}~,~(ab:c)\equiv (a\cdot c)(b\cdot c)-c^2(a\cdot b)~.
\end{eqnarray}
Finally, in the $\nu'$-formalism I use the following representation of the dot products:
\begin{eqnarray}
&&p^2=m^2~,~p\cdot p_e=m E_e~,~p\cdot p_\nu'=mE_\nu'~,~p_e^2=m_e^2~,~p_e\cdot p_\nu'=E_e E_\nu'-|\vec{p}_e|\pnu c'~,\nonumber\\
&&p_\nu^{\prime 2}=E_\nu^{\prime 2}-\pnu^2~,
\end{eqnarray}
whereas, in the recoil formalism I use:	
\begin{eqnarray}
&&p^2=m^2~,~p\cdot p_e=\frac{m^2y}{2}~,~p\cdot p_\nu'=\frac{m^2}{2}(2-y-z)~,~p_e^2=m^2r_e~,\nonumber\\
&&p_e\cdot p_\nu'=\frac{m^2}{2}(1+r_f-r_e-x-z)~,~p_\nu^{\prime 2}=m^2x~.
\end{eqnarray}

\section{\label{sec:spin}$p_\nu$- and $k$-integral with the spin vector}

In this work I allow the parent nucleus to be polarized. This introduces the spin vector $s^\mu$ which can appear at most linearly in the squared amplitude. So, in doing the $p_\nu$- and $k$-integrations, one may also encounter integrals of the form
\begin{equation}
\int\frac{d^3k}{(2\pi)^32E_k}\frac{d^3p_\nu}{(2\pi)^32E_\nu}(2\pi)^4\delta^{(4)}(p_\nu'-p_\nu-k)\frac{s\cdot k}{(p\cdot k)^i(p_e\cdot k)^j}
\end{equation}
in addition to those in Appendix~\ref{sec:Imn}. To evaluate it, I first parameterize:
\begin{equation}
I_{i,j}^\mu\equiv \int\frac{d^3k}{(2\pi)^32E_k}\frac{d^3p_\nu}{(2\pi)^32E_\nu}(2\pi)^4\delta^{(4)}(p_\nu'-p_\nu-k)\frac{k^\mu}{(p\cdot k)^i(p_e\cdot k)^j}=c_1 p^\mu+c_2p_e^\mu+c_3p_\nu^{\prime \mu}~,\label{eq:Imumn}
\end{equation}	
and evaluate the coefficients $c_{1,2,3}$. Contracting Eq.\eqref{eq:Imumn} with $p_\mu$, $p_{e\mu}$ and $(p_\nu')_\mu$ gives rise to the following matrix equation:
\begin{equation}
\left(\begin{array}{c}
c_{1}\\
c_{2}\\
c_{3}
\end{array}\right)=M^{-1}\left(\begin{array}{c}
I_{i-1,j}\\
I_{i,j-1}\\
\frac{p_{\nu}^{\prime2}}{2}I_{i,j}
\end{array}\right)~,~M= \left(\begin{array}{ccc}
m_{N}^{2} & p\cdot p_{e} & p\cdot p_{\nu}'\\
p\cdot p_{e} & m_{e}^{2} & p_{e}\cdot p_{\nu}'\\
p\cdot p_{\nu}' & p_{e}\cdot p_{\nu}' & p_{\nu}^{\prime2}
\end{array}\right)~.
\end{equation}
Plugging this back to Eq.\eqref{eq:Imumn} and contracting to $s_{\mu}$ gives:
\begin{eqnarray}
s_{\mu}I_{i,j}^\mu&=&\left[(M^{-1})_{21}I_{i-1,j}+(M^{-1})_{22}I_{i,j-1}+\frac{p_\nu^{\prime 2}}{2}(M^{-1})_{23}I_{i,j}\right]s\cdot p_e\nonumber\\
&&+\left[(M^{-1})_{31}I_{i-1,j}+(M^{-1})_{32}I_{i,j-1}+\frac{p_\nu^{\prime 2}}{2}(M^{-1})_{33}I_{i,j}\right]s\cdot p_\nu'~.
\end{eqnarray} 
So, the new integral is expressible in terms of the old integrals $I_{i,j}$. In fact, this result can be conveniently implemented by making the following replacement at the squared amplitude:
\begin{eqnarray}
s\cdot k&\rightarrow&\left[(M^{-1})_{21}s\cdot p_e+(M^{-1})_{31}s\cdot p_\nu'\right]p\cdot k+\left[(M^{-1})_{22}s\cdot p_e+(M^{-1})_{32}s\cdot p_\nu'\right]p_e\cdot k\nonumber\\
&&+\left[(M^{-1})_{23}s\cdot p_e+(M^{-1})_{33}s\cdot p_\nu'\right]p_\nu^{\prime 2}/2~.\label{eq:skreplace}
\end{eqnarray}
	
\section{\label{sec:Cij}The coefficients $C_{ij}^X$}	

Here I provide the analytic expressions of the coefficients $C_{ij}^X$ defined in Eq.\eqref{eq:Mreg2replace}.

\subsection{$C_{ij}^0$}

The non-zero coefficients are:
\begin{eqnarray}
C_{00}^0&=&\frac{16\pi(\lambda^2+1)}{E_eE_\nu^{\prime 2}}\nonumber\\
C_{10}^0&=&\frac{8\pi m(4E_e-(1+3\lambda^2)E_\nu')}{E_eE_\nu^{\prime 2}}\nonumber\\
C_{01}^0&=&-\frac{4\pi}{E_e E_\nu^{\prime 2}}\left[(\lambda^2-1)(2|\vec{p}_e| \pnu c'+\pnu^2-E_\nu^{\prime 2}-2m_e^2)+8E_e^2(\lambda^2+1)-2E_eE_\nu'(5\lambda^2+3)\right]\nonumber\\
C_{02}^0&=&\frac{4\pi(\lambda^2-1)m_e^2(\pnu^2-E_\nu^{\prime 2})}{E_eE_\nu^{\prime 2}}\nonumber\\
C_{11}^0&=&\frac{4\pi(\lambda^2-1)m(E_\nu^{\prime 2}-\pnu^2)}{E_\nu^{\prime 2}}\nonumber\\
C_{-11}^{0}&=&-\frac{16\pi(\lambda^2+1)(2E_e-E_\nu')}{E_eE_\nu^{\prime 2}m}\nonumber\\
C_{2-1}^{0}&=&\frac{8\pi(\lambda^2-1)m^2}{E_eE_\nu^{\prime 2}}\nonumber\\
C_{-12}^0&=&\frac{16\pi(\lambda^2+1)m_e^2(E_e-E_\nu')}{E_e E_\nu^{\prime 2}m}\nonumber\\
C_{-21}^0&=&-\frac{16\pi(\lambda^2+1)}{E_eE_\nu^{\prime 2}m^2}\nonumber\\
C_{-22}^0&=&\frac{16\pi(\lambda^2+1)m_e^2}{E_eE_\nu^{\prime 2}m^2}
\end{eqnarray}	
It is worth noticing that $C_{ij}^0$ is independent of $\zeta$, i.e. they are the same for $\beta^\pm$ decay. This is not the case for the spin-dependent coefficients $C_{ij}^\beta$ and $C_{ij}^\nu$. 

\subsection{$C_{ij}^\beta$}

The non-zero coefficients are:
\begin{eqnarray}
C_{00}^\beta&=&-\frac{16\pi\lambda\left[\pnu(-2(\lambda\zeta+2)E_e+(-\lambda\zeta+1)E_\nu')+|\vec{p}_e|c'E_\nu'(\lambda\zeta+1)\right]}{|\vec{p}_e|^2\pnu(c^{\prime 2}-1)E_\nu^{\prime 2}}\nonumber\\
C_{-10}^\beta&=&\frac{32\pi\lambda}{|\vec{p}_e|^2(c^{\prime 2}-1)E_\nu^{\prime 2}m}\nonumber\\
C_{10}^\beta&=&\frac{8\pi\lambda m}{|\vec{p}_e|^2\pnu(c^{\prime 2}-1)E_\nu^{\prime 2}}\left[2|\vec{p}_e|^2\pnu c^{\prime 2}(\lambda\zeta-1)-|\vec{p}_e|c'(\lambda\zeta+1)(\pnu^2+E_\nu'(2E_e-E_\nu'))\right.\nonumber\\
&&\left.+2\pnu(2E_e^2(\lambda\zeta+2)+(E_eE_\nu'+m_e^2)(\lambda\zeta-1))\right]\nonumber\\
C_{01}^\beta&=&\frac{-8\pi\lambda }{|\vec{p}_e|^2\pnu(c^{\prime 2}-1)E_\nu^{\prime 2}}\left[2|\vec{p}_e|^2\pnu c^{\prime 2}(\lambda\zeta-1)(2E_e-E_\nu')-|\vec{p}_e|\pnu^2 c'(E_e(\lambda\zeta+3)+E_\nu'(\lambda\zeta-1))\right.\nonumber\\
&&-|\vec{p}_e|E_\nu'c'(4E_e^2(\lambda\zeta+1)+E_eE_\nu'(\lambda\zeta-5)-E_\nu^{\prime 2}(\lambda\zeta-1))\nonumber\\
&&\left.+2\pnu E_e(4E_e^2+2E_eE_\nu'(\lambda\zeta-1)+(3\lambda\zeta-1)m_e^2)\right]\nonumber\\
C_{20}^\beta&=&\frac{-8\pi c'E_e(\lambda\zeta+1)\lambda m^2(\pnu^2-E_\nu^{\prime 2})}{|\vec{p}_e|\pnu(c^{\prime 2}-1)E_\nu^{\prime 2}}\nonumber\\
C_{02}^\beta&=&\frac{-8\pi c'\lambda m_e^2(\pnu^2-E_\nu^{\prime 2})(E_e(\lambda\zeta+1)+E_\nu'(\lambda\zeta-1))}{|\vec{p}_e|\pnu(c^{\prime 2}-1)E_\nu^{\prime 2}}\nonumber\\
C_{11}^\beta&=&\frac{8\pi c'E_e\lambda m(\pnu^2-E_\nu^{\prime 2})(2E_e(\lambda\zeta+1)+E_\nu'(\lambda\zeta-1))}{|\vec{p}_e|\pnu (c^{\prime 2}-1)E_\nu^{\prime 2}}\nonumber\\
C_{1-1}^\beta&=&-\frac{16\pi(\lambda\zeta+1)\lambda m}{|\vec{p}_e|^2(c^{\prime 2}-1)E_\nu^{\prime 2}}\nonumber\\
C_{-11}^\beta&=&\frac{-16\pi\lambda}{|\vec{p}_e|^2\pnu(c^{\prime 2}-1)E_\nu^{\prime 2}m}\left[|\vec{p}_e|^2\pnu c^{\prime 2}(\lambda\zeta-1)-|\vec{p}_e|\pnu^2c'\right.\nonumber\\
&&\left.-|\vec{p}_e|E_\nu'c'(E_e(\lambda\zeta+3)+E_\nu'(\lambda\zeta-2))+\pnu(4E_e^2+E_eE_\nu'(\lambda\zeta-1)+(\lambda\zeta+1)m_e^2)\right]\nonumber\\
C_{2-1}^\beta&=&-\frac{16\pi E_e(\lambda\zeta+1)\lambda m^2}{|\vec{p}_e|^2(c^{\prime 2}-1)E_\nu^{\prime 2}}\nonumber\\
C_{-12}^\beta&=&\frac{16\pi\lambda m_e^2}{|\vec{p}_e|^2\pnu(c^{\prime 2}-1)E_\nu^{\prime 2}m}\left[|\vec{p}_e|^2\pnu c^{\prime 2}(\lambda\zeta-1)-|\vec{p}_e|\pnu^2c'\right.\nonumber\\
&&\left.-|\vec{p}_e|E_\nu'c'(E_e(\lambda\zeta+1)+E_\nu'(\lambda\zeta-2))+\pnu(2E_e^2+(E_eE_\nu'+m_e^2)(\lambda\zeta-1))\right]\nonumber\\
C_{-21}^\beta&=&\frac{32\pi\lambda(|\vec{p}_e|E_\nu'c'-\pnu E_e)}{|\vec{p}_e|^2\pnu(c^{\prime 2}-1)E_\nu^{\prime 2}m^2}\nonumber\\
C_{-22}^\beta&=&\frac{32\pi\lambda m_e^2(\pnu E_e-|\vec{p}_e|E_\nu'c')}{|\vec{p}_e|^2\pnu(c^{\prime 2}-1)E_\nu^{\prime 2}m^2}
\end{eqnarray}

\subsection{$C_{ij}^\nu$}

The non-zero coefficients are:
\begin{eqnarray}
C_{00}^\nu&=&-\frac{16\pi\lambda\left[\pnu c'(2(\lambda\zeta+2)E_e+(\lambda\zeta-1)E_\nu')-|\vec{p}_e|E_\nu'(\lambda\zeta+1)\right]}{|\vec{p}_e|\pnu(c^{\prime 2}-1)E_eE_\nu^{\prime 2}}\nonumber\\
C_{-10}^\nu&=&-\frac{32\pi\lambda c'}{|\vec{p}_e|(c^{\prime 2}-1)E_eE_\nu^{\prime 2}m}\nonumber\\
C_{10}^\nu&=&\frac{8\pi\lambda m}{|\vec{p}_e|\pnu(c^{\prime 2}-1)E_eE_\nu^{\prime 2}}\left[|\vec{p}_e|(\lambda\zeta+1)(\pnu^2(2c^{\prime 2}-1)+E_\nu'(2E_e-E_\nu'))\right.\nonumber\\
&&\left.-2\pnu E_e c'(3E_e(\lambda\zeta+1)+E_\nu'(\lambda\zeta-1))\right]\nonumber\\
C_{01}^\nu&=&\frac{-8\pi\lambda }{|\vec{p}_e|\pnu(c^{\prime 2}-1)E_eE_\nu^{\prime 2}}\left[|\vec{p}_e|\pnu^2(E_e(4c^{\prime 2}(\lambda\zeta+1)-3\lambda\zeta-1)+E_\nu'(\lambda\zeta-1))\right.\nonumber\\
&&+|\vec{p}_e|E_\nu'(4E_e^2(\lambda\zeta+1)+E_e E_\nu'(\lambda\zeta-5)-E_\nu^{\prime 2}(\lambda\zeta-1))\nonumber\\
&&\left.-2\pnu c'(E_e(2E_e^2+m_e^2)(\lambda\zeta+1)+E_\nu'(E_e^2+m_e^2)(\lambda\zeta-1))\right]\nonumber\\
C_{20}^\nu&=&\frac{8\pi (\lambda\zeta+1)\lambda m^2(\pnu^2-E_\nu^{\prime 2})}{\pnu(c^{\prime 2}-1)E_\nu^{\prime 2}}\nonumber\\
C_{02}^\nu&=&\frac{8\pi\lambda m_e^2(\pnu^2-E_\nu^{\prime 2})(E_e(\lambda\zeta+1)+E_\nu'(\lambda\zeta-1))}{\pnu(c^{\prime 2}-1)E_eE_\nu^{\prime 2}}\nonumber\\
C_{11}^\nu&=&-\frac{8\pi \lambda m(\pnu^2-E_\nu^{\prime 2})(2E_e(\lambda\zeta+1)+E_\nu'(\lambda\zeta-1))}{\pnu (c^{\prime 2}-1)E_\nu^{\prime 2}}\nonumber\\
C_{1-1}^\nu&=&\frac{16\pi c'(\lambda\zeta+1)\lambda m}{|\vec{p}_e|(c^{\prime 2}-1)E_eE_\nu^{\prime 2}}\nonumber\\
C_{-11}^\nu&=&\frac{-16\pi\lambda}{|\vec{p}_e|\pnu(c^{\prime 2}-1)E_eE_\nu^{\prime 2}m}\left[|\vec{p}_e|\pnu^2(c^{\prime 2}(\lambda\zeta+1)-\lambda\zeta)+|\vec{p}_e|E_\nu'(E_e(\lambda\zeta+3)+E_\nu'(\lambda\zeta-2))\right.\nonumber\\
&&\left.-\pnu c'(E_e^2(\lambda\zeta+3)+E_eE_\nu'(\lambda\zeta-1)+2m_e^2)\right]\nonumber\\
C_{2-1}^\nu&=&\frac{16\pi c'(\lambda\zeta+1)\lambda m^2}{|\vec{p}_e|(c^{\prime 2}-1)E_\nu^{\prime 2}}\nonumber\\
C_{-12}^\nu&=&\frac{16\pi\lambda m_e^2}{|\vec{p}_e|\pnu(c^{\prime 2}-1)E_eE_\nu^{\prime 2}m}\left[|\vec{p}_e|\pnu^2(c^{\prime 2}(\lambda\zeta+1)-\lambda\zeta)+|\vec{p}_e|E_\nu'(E_e(\lambda\zeta+1)+E_\nu'(\lambda\zeta-2))\right.\nonumber\\
&&\left.-\pnu c'E_e(E_e(\lambda\zeta+1)+E_\nu'(\lambda\zeta-1))\right]\nonumber\\
C_{-21}^\nu&=&\frac{32\pi\lambda(\pnu E_ec'-|\vec{p}_e|E_\nu')}{|\vec{p}_e|\pnu(c^{\prime 2}-1)E_eE_\nu^{\prime 2}m^2}\nonumber\\
C_{-22}^\nu&=&\frac{32\pi\lambda m_e^2(|\vec{p}_e|E_\nu'-\pnu E_e c')}{|\vec{p}_e|\pnu(c^{\prime 2}-1)E_eE_\nu^{\prime 2}m^2}
\end{eqnarray}
	
\section{\label{sec:dij}The coefficients $d_{ij}$}		

Here I provide the analytic expressions of the non-zero coefficients $d_{ij}$ defined in Eq.\eqref{eq:recoilexpand} in the recoil formalism.

\begin{eqnarray}
d_{00}&=&32\pi(\lambda^2+1)\nonumber\\
d_{10}&=&8\pi m^2\left[(3\lambda^2+5)y+(3\lambda^2+1)(z-2)\right]\nonumber\\
d_{01}&=&8\pi m^2\left[(\lambda^2-1)(r_e+r_f-z+1)-(\lambda^2+1)(4y^2+2yz-4y)\right]\nonumber\\
d_{02}&=&-8\pi m^4(\lambda^2-1) r_e x\nonumber\\
d_{11}&=&4\pi m^4(\lambda^2-1)xy\nonumber\\
d_{-11}&=&-16\pi(\lambda^2+1)(3y+z-2)\nonumber\\
d_{2-1}&=&16\pi m^2(\lambda^2-1)\nonumber\\
d_{-12}&=&16\pi m^2(\lambda^2+1)r_e(2y+z-2)\nonumber\\
d_{-21}&=&-\frac{32\pi}{m^2}(\lambda^2+1)\nonumber\\
d_{-22}&=&32\pi(\lambda^2+1)r_e
\end{eqnarray} 
	
\end{appendix}

\bibliography{ref}

\end{document}